\documentclass[twocolumn,aps,showpacs,prb,tightenlines,amsmath,amssymb]{revtex4}
\usepackage{graphicx}
\usepackage{amssymb}
\usepackage{slashed}
\usepackage{dcolumn}
\usepackage{amsmath}
\usepackage{bm}
\usepackage{colordvi}
\usepackage{mathrsfs}
\makeatletter

\newcommand{\Rmnum}[1]{\expandafter\@slowromancap\romannumeral #1@}
\makeatother

\begin{document}

\title{Optical response of Higgs mode in superconductors at clean limit}   

\author{F. Yang}
\email{yfgq@mail.ustc.edu.cn.}

\affiliation{Hefei National Laboratory for Physical Sciences at
Microscale, Department of Physics, and CAS Key Laboratory of Strongly-Coupled
Quantum Matter Physics, University of Science and Technology of China, Hefei,
Anhui, 230026, China}

\author{M. W. Wu}
\email{mwwu@ustc.edu.cn.}

\affiliation{Hefei National Laboratory for Physical Sciences at
Microscale, Department of Physics, and CAS Key Laboratory of Strongly-Coupled
Quantum Matter Physics, University of Science and Technology of China, Hefei,
Anhui, 230026, China}

\date{\today}

\begin{abstract}

  The phenomenological Ginzburg-Landau theory and the charge conservation directly lead to the finite Higgs-mode generation and vanishing charge-density fluctuation in the second-order optical response of superconductors at clean limit. Nevertheless, recent microscopic theoretical studies of the second-order optical response, apart from the one through the gauge-invariant kinetic equation [Yang and Wu, Phys. Rev. B {\bf 100}, 104513 (2019)], have derived a vanishing Higgs-mode generation but finite charge-density fluctuation at clean limit. We resolve this controversy by re-examining the previous derivations with the vector potential alone within the path-integral and Eilenberger-equation approaches, and show that both previous derivations contain flaws. After fixing these flaws, a finite Higgs-mode generation through the drive effect of vector potential is derived at clean limit, exactly recovering the previous result from the gauge-invariant kinetic equation as well as Ginzburg-Landau theory. By further extending the path-integral approach to include electromagnetic effects from the scalar potential and phase mode, in the second-order response, a finite contribution from the drive effect of scalar potential to the Higgs-mode generation at clean limit as well as the vanishing charge-density fluctuation are derived, also recovering the results from the gauge-invariant kinetic equation. Particularly, we show that the phase mode is excited in the second-order response, and exactly cancels the previously reported unphysical excitation of the charge-density fluctuation, guaranteeing the charge conservation.

\end{abstract}


\maketitle 

\section{Introduction}

Recently, the Higgs mode in superconductors has attracted extensive experimental and theoretical interest. Specifically, the angular and radial excitations that emerge in the Mexican-hat free energy of superconductors, describe the phase and amplitude fluctuations of the superconducting order parameter $\Delta$\cite{Am0}, respectively. The gapless phase mode $\delta\theta$, referred to as Nambu-Goldstone mode\cite{gi0,AK,Gm1,Gm2,Ba0,pm0,pi1,pm1,pm2,pi4,gi1,Ba9,Ba10,pm5,AK2}, corresponds to gapless Goldstone boson due to the spontaneous breaking of continuous $U(1)$ symmetry\cite{Gm1,Gm2}. Whereas the amplitude mode, exhibiting a gapful energy spectrum $\omega_H$, is referred to as the Higgs mode\cite{OD1,OD2,OD3,pm5,Am0,Am5,Am6,Am12,AK2}, because of the similarity to Higgs boson in the field theory\cite{Higgs1,Higgs2,Higgs3}. Early theoretical works in conventional $s$-wave superconductors have reported $\omega_H=2|\Delta_0|$ at long-wave limit\cite{pm5,Am0,OD1,OD2,OD3,Am6,AK2}, with $|\Delta_0|$ being the superconducting gap. Nevertheless, being charge neutral and spinless, the Higgs mode has long been experimentally elusive. Until recently, thanks to the advanced technique in ultrafast nonlinear optics, it is experimentally realized\cite{NL7,NL8,NL9,NL10,NL11,DHM2,DHM3} that an intense terahertz optical field can excite the fluctuation $\delta\rho_s$ of the superfluid density $\rho_s$ in the second-order optical response, which manifests itself in the third-harmonic current. This fluctuation is attributed to the excitation of the Higgs mode, due to an observed resonance when twice of the optical frequency  is tuned at $\omega_H$\cite{NL8,NL9,NL10}. Inspired by the experimental finding, a great deal of theoretical efforts have been devoted to the nonlinear optical response of superconductors. However, rather than straightening  out the situation, these theoretical descriptions make the understanding of the existing and growing experimental findings muddled.  

Specifically, early theoretical studies of the nonlinear optical response in superconductors have used the Bloch\cite{Am1,Am2,Am7,Am9,Am11,Am14,Am15,NL7,NL8,NL9,NL10,NL11} or Liouville\cite{Am3,Am4,Am8,Am10,Am16} equation derived in Anderson pseudospin picture\cite{As}, with the vector potential ${\bf A}$ alone. The second-order light-matter interaction $H_{p}=e^2A^2\tau_3/(2m)$ naturally emerges in these descriptions\cite{Am1,Am2,Am3,Am4,Am7,Am8,Am9,Am10,Am11,Am14,NL7,NL8,NL9,NL10,NL11} as a pseudo field along $z$ direction, with $\tau_{i}$ being the Pauli matrices in Nambu space. This interaction can pump the quasiparticle correlation (i.e., pseudospin precession) and then causes the fluctuation of the order parameter, which was directly considered as the Higgs mode to explain the experimental findings\cite{Am1,Am2,Am3,Am4,Am7,Am8,Am9,Am10,Am11,Am14,NL7,NL8,NL9,NL10,NL11}. Nevertheless, a latter symmetry analysis in Anderson pseudospin picture finds a vanishing (finite) correlation between amplitude (phase) mode and external pseudo field\cite{symmetry}. This implies that the theoretically obtained order-parameter fluctuation in Bloch or Liouville equation is a phase fluctuation rather than the claimed amplitude one\cite{GIKE2}. Moreover, with the isotropic pump effect alone, the Bloch or Liouville equation in the literature\cite{Am1,Am2,Am3,Am4,Am7,Am8,Am9,Am10,Am11,Am14,NL7,NL8,NL9,NL10,NL11} fails to drive the optical current and hence derive the superfluid density, since no drive effect (i.e., linear light-matter interaction) is included. 

By applying the standard path-integral approach with the vector potential alone, Cea {\em et al.}\cite{Cea1,Cea2,Cea3} further considered the linear-order light-matter interaction $H_d={\hat{\bf p}}\cdot{e}{\bf A}/m$ (i.e., drive effect from vector potential\cite{G1,GIKE1}, with ${\hat{\bf p}}$ being the momentum operator), in addition to second-order one $H_p$ (pump effect). In the second-order response, they found\cite{Cea1,Cea2,Cea3} that neither drive nor pump effects can excite the Higgs mode $\delta|\Delta|$. Starting from this theoretical investigation, it is believed afterwards that the Higgs-mode generation is zero at clean limit. Meanwhile, Cea {\em et al.}\cite{Cea1,Cea2,Cea3} found that the pump effect $H_p$ can cause the fluctuation $\delta{n}$ of the charge density $n$. As the superfluid density $\rho_s$ is proportional to $n|\Delta|^2$\cite{G1}, it is therefore speculated\cite{Cea1,Cea2,Cea3} that the experimentally observed $\delta\rho_s$ is attributed to the charge-density fluctuation $\delta{n}$ rather than the Higgs mode $\delta|\Delta|$. Several polarization-resolved measurements were performed afterwards\cite{NL10,NL11,DHM2,DHM3,FHM}, as the theoretically predicted signal of the Higgs mode (charge-density fluctuation) is isotropic (anisotropic)\cite{Cea1}. However, an isotropic optical response is experimentally observed\cite{NL10,NL11,DHM2,DHM3}, giving firm evidence to support the previous observation of the Higgs mode. Recent theoretical attention then tends to focus on and emphasize the important role of the impurity scattering to mediate the Higgs-mode generation\cite{Am16,ImR1,ImR2,ImR3,Silaev}. To handle the microscopic scattering seriously, Silaev\cite{Silaev,Silaev0} used the Eilenberger equation\cite{Eilen,Ba20,Eilen1} that only involves the drive effect $H_d$ by vector potential. In the second-order response, he also found a vanishing Higgs-mode excitation at clean limit, but derived a finite one in dirty case to dominate over the charge-density fluctuation\cite{Silaev}.

Through a gauge-invariant kinetic equation (GIKE) approach with complete electromagnetic effect\cite{GIKE1,GIKE2,GIKE3,GIKE4}, our recent study\cite{GIKE2} that calculates the amplitude and phase modes on an equal footing obtained totally different results in the second-order optical response at clean limit:  a vanishing charge-density fluctuation, and a finite Higgs-mode generation, contributed by the drive effect $H_d$ of vector potential. Physically, both results can be understood as follows. Firstly, it is well known from the symmetry analysis that there is no second-order current {\small $j^{(2)}$} in systems with the inversion symmetry. Then, from the charge conservation {\small $\partial_te\delta{n}+{\bm \nabla_R}\cdot{\bf j}=0$} in the second-order regime ({\small $2{\Omega}e{\delta}n^{(2)}+2{\bf q}\cdot{\bf j}^{(2)}=0$}, with $\Omega$ and ${\bf q}$ being the optical frequency and momentum, respectively), the second-order charge-density fluctuation {\small $e{\delta}n^{(2)}$} is forbidden. Secondly, according to the Ginzburg-Landau theory, the general superconducting Lagrangian at clean limit reads\cite{Am6,PI2GL,PI2GL2,G1}
\begin{equation}
\mathscr{L}\!=\!\frac{\gamma_L|i\partial_t\Psi|^2}{2}\!-\!\Big(\alpha_L|\Psi|^2\!+\!\frac{\beta_L|\Psi|^4}{2}\!+\!\frac{\lambda_L|({\bm \nabla}\!-\!2ie{\bf A})\Psi|^2}{4m}\Big),  \label{GLLE}
\end{equation}
with $\Psi$ denoting the order parameter and $\gamma_L$, $\alpha_L$, $\beta_L$ as well as $\lambda_L$ representing the Landau parameters. From the Lagrangian above, the equilibrium order parameter $\Psi_0=\sqrt{-\alpha_L/\beta_L}$. Whereas by only considering the amplitude fluctuation $\delta|\Psi|$ with $\Psi=\Psi_0+\delta|\Psi|$, one can directly obtain its equation of motion:
\begin{equation}\label{GL-HME}
\big[(2|\Psi_0|)^2+\partial_{t}^2\big]\delta|\Psi|=-\frac{\lambda_L}{\beta_L}\frac{e^2A^2}{m}2\Psi_0,  
\end{equation}
where {\small $\beta_L=\gamma_L=\frac{7R(3)}{8(\pi{T})^2}$} and {\small $\lambda_L=\frac{k_F^2}{3m}\frac{7R(3)}{4(\pi{T})^2}$} are used for conventional $s$-wave superconductors\cite{G1,PI2GL}, with $T$ denoting temperature and $R(x)$ being the Riemann zeta function. Then, from above equation, one can immediately find the Higgs-mode energy spectrum $\omega_H=2|\Psi_0|$ on the left-hand side of the equation, and in particular, a finite second-order response of Higgs mode at clean limit on the right-hand side of the equation. Moreover, it is established in the derivation of the Ginzburg-Landau equation from the basic Gorkov equation\cite{G1} that the kinetic term [i.e., the last term in Eq.~(\ref{GLLE})] is solely attributed to the drive effect $H_d$, implying a finite (zero) contribution from the drive (pump) effect to Higgs-mode excitation.

Consequently, in the second-order response at clean limit, the finite Higgs-mode excitation and vanishing charge-density fluctuation derived from GIKE\cite{GIKE2} and justified by the phenomenological Ginzburg-Landau theory and the charge conservation mentioned above, pose a sharp contrast to the previous derivations from the path-integral approach by Cea {\em et al.}\cite{Cea1,Cea2,Cea3} and Eilenberger equation by Silaev\cite{Silaev} where zero Higgs-mode generation and finite charge-density fluctuation are obtained. Moreover, it also becomes particularly bizarre that both path-integral approach\cite{PI2GL} and Eilenberger equation\cite{Ba20} can recover the Ginzburg-Landau equation, but obtained a zero Higgs-mode generation\cite{Cea1,Cea2,Cea3,Silaev} which holds against the Ginzburg-Landau theory. 

In the present work, we resolve this controversy by re-examining the previous derivations with the vector potential alone within the path-integral approach by Cea {\em et al.}\cite{Cea1,Cea2,Cea3} and Eilenberger equation by Silaev\cite{Silaev}. While we have successfully recovered their results, it is found that both derivations contain flaws. In the path-integral approach, after integration over the Fermi field, the coupling of the Higgs mode to the pump effect $H_p$ and the second order of the drive effect $H_d$ emerge in the second- and third-order perturbation expansions of the action, respectively. Nevertheless, in Refs.~\onlinecite{Cea1,Cea2,Cea3}, only the second-order expansion is kept, leading to zero Higgs-mode generation because of the vanishing correlation between amplitude mode and pump effect\cite{symmetry}. The essential third-order expansion, which is related to the Ginzburg-Landau kinetic term\cite{PI2GL} and hence finite Higgs-mode generation, is excessively overlooked.
In Ref.~\onlinecite{Silaev} within the Eilenberger equation, in the summation over the Fermion Matsubara frequency, the involved continuous optical frequency $\Omega$ over $\pi{T}$ is considered as discrete even integer, leading to vanishing amplitude-response coefficient. We prove that after fixing these flaws, in the second-order optical response at clean limit, one can find a finite Higgs-mode generation contributed by the drive effect $H_d$, exactly recovering the results from GIKE\cite{GIKE2} as well as the Ginzburg-Landau theory.

Generally, according to the gauge structure in superconductors first revealed by Nambu\cite{gi0}, among the scalar potential $\phi$, vector potential ${\bf A}$ as well as the superconducting-phase-related effective electromagnetic potential {\small $\partial_{\mu}\delta\theta$}, one can not choose two quantities simultaneously to be zero in superconductors, e.g., considering the vector potential alone.  Therefore, we further extend the previous path-integral approach to include the electromagnetic effects from the scalar potential and phase mode, which have been overlooked in the previous theoretical descriptions\cite{Am1,Am2,Am3,Am4,Am7,Am8,Am9,Am10,Am11,Am14,NL7,NL8,NL9,NL10,NL11,Cea1,Cea2,Cea3,Silaev}. Then, in the second-order response at clean limit, the finite contribution in the Higgs-mode generation from the scalar potential as well as the vanishing charge-density fluctuation, both of which have previously been obtained from GIKE\cite{GIKE2}, are recovered. On one hand, differing from the contribution of the vector potential that emerges only at finite temperature\cite{DS1}, the one of the scalar potential is finite upon cooling to zero temperature, and is essential because of the gauge structure in superconductors\cite{gi0}. On the other hand, we show that a  spatially uniform phase mode is generated in the second-order response, and exactly cancels the unphysical excitation of the charge-density fluctuation reported by Cea {\em et al.}\cite{Cea1,Cea2,Cea3}, guaranteeing the charge conservation. Consequently, the present study arrives at unified conclusions about the finite Higgs-mode generation and vanishing charge-density fluctuation in the second-order optical response at clean limit, in consistency with the phenomenological Ginzburg-Landau theory and the charge conservation, and hence, can help understanding the experimental findings. Furthermore,  a disscussion about the application of Matsubara formalism in the derivation of superconducting gap dynamics is presented.

\section{Hamiltonian}
\label{sec-H}

We begin with the Bogoliubov-de Gennes Hamiltonian of the conventional $s$-wave superconducting states in the presence of the electromagnetic potential $A_{\mu}=(\phi,{\bf A})$\cite{G1}: 
\begin{equation}
H={\int}{d{\bf x}}\psi^{\dagger}(x)[(\xi_{{\hat{\bf p}}-e{\bf
    A}\tau_3}+e\phi+\mu_H)\tau_3+{\hat \Delta}(x)]\psi(x),\label{BdG}
\end{equation}
where $\psi(x)=[\psi_{\uparrow}(x),\psi^{\dagger}_{\downarrow}(x)]^T$ is the field operator in the Nambu space; $x=(x_0,{\bf x})$ denotes the space-time vector; $\xi_{\hat{\bf p}}={{\bf{\hat p}}^2}/({2m})-\mu$ with $m$ and $\mu$ being the effective mass and chemical potential; the momentum operator ${\hat {\bf p}}=-i\hbar{\bm \nabla}$; $\mu_H(x)=\sum_{x'}V(x-x')\delta{n}(x')$ denotes the Hartree field, which is equivalent to the Poisson equation and characterizes the induced scalar potential by density fluctuation $\delta{n}(x')$; $V(x-x')$ represents the Coulomb potential;  ${\hat \Delta}(x)=\Delta(x)\tau_++{\Delta}^*(x)\tau_-$, while considering the phase and amplitude modes, the order parameter reads $\Delta(x)=[\Delta_{0}+\delta|\Delta|(x)]e^{i\delta\theta(x)}$. 

It is noted that the phase mode $\delta\theta$ in Eq.~(\ref{BdG}) can be effectively removed by a unitary transformation:
\begin{equation}\label{utrp}
\psi(x){\rightarrow}e^{i\tau_3\delta\theta(x)/2}\psi(x),
\end{equation}
and then, one has\cite{gi0,AK}
\begin{equation}
  H=\!\!\!{\int}{d{\bf x}}\psi^{\dagger}(x)(H_0+H_{\rm LM}+\delta|\Delta|\tau_1)\psi(x),
\end{equation}
with the free BCS Hamiltonian 
\begin{equation}
H_0=\xi_{\bf {\hat p}}\tau_3+\Delta_0\tau_1,  
\end{equation}  
and the light-matter interaction
\begin{equation}
  H_{\rm LM}=\frac{{\bf p}_s\cdot{\hat{\bf p}}}{m}+\frac{p_s^2}{2m}\tau_3+\mu_{\rm eff}\tau_3. \label{LM}
\end{equation}
Here, the gauge-invariant superconducting momentum {\small ${\bf p}_s={\bm \nabla}_{\bf x}\delta\theta/2-e{\bf A}$} and effective field {\small $\mu_{\rm eff}=e\phi+\partial_{x_0}\delta\theta/2+\mu_{H}$}\cite{gi0,AK}. Then, it is clearly seen that the phase mode provides an effective electromagnetic potential {\small $eA^{\rm eff}_{\mu}=(\partial_{x_0}\delta\theta/2,-{\bm \nabla_{\bf x}}\delta\theta/2)$}, in consistency with the gauge structure in superconductors first revealed by Nambu\cite{gi0}:
\begin{eqnarray}
eA_{\mu}&\rightarrow&eA_{\mu}-\partial_{\mu}\chi, \label{gaugestructure1}\\
\label{gaugestructure2}
\delta\theta&\rightarrow&\delta\theta+2\chi.
\end{eqnarray}
Here, $\partial_{\mu}=(\partial_{x_0},-{\bm \nabla}_{\bf x})$. It is noted that the gauge invariance is essential for the theoretical descriptions, since it guarantees the charge conservation, as first proved by Nambu via the generalized Ward's identity\cite{gi0,Ba0}.

By assuming the electromagnetic potential {\small ${\bf A}(x)={\bf A}_0e^{i\Omega{x_0}-i{\bf q}\cdot{\bf x}}$} and {\small $\phi(x)=\phi_0({\bf x})e^{i\Omega{x_0}-i{\bf q}\cdot{\bf x}}$} with {\small ${\phi_0({\bf x})}={\bar \phi_0}+{\bf E}_{\phi}\cdot{\bf x}$} and {\small $E_{\phi}$} being the transverse field, one has 
\begin{eqnarray}
e\delta{n}&=&e{\delta}n^{(1)}e^{i\Omega{x_0}-i{\bf q}\cdot{\bf 
  x}}+{\delta}n^{(2)}e^{2i\Omega{x_0}-2i{\bf q}\cdot{\bf x}},  \\
\delta\theta&=&\delta\theta^{(1)}e^{i\Omega{x_0}-i{\bf q}\cdot{\bf 
  x}}+\delta\theta^{(2)}e^{2i\Omega{x_0}-2i{\bf q}\cdot{\bf 
  x}},\\
\delta|\Delta|&=&\delta|\Delta|^{(1)}e^{i\Omega{x_0}-i{\bf q}\cdot{\bf 
  x}}+\delta|\Delta|^{(2)}e^{2i\Omega{x_0}-2i{\bf q}\cdot{\bf 
  x}},
\end{eqnarray} 
where $e{\delta}n^{(l)}$ as well as $\delta\theta^{(l)}$ and $\delta|\Delta|^{(l)}$ denote the $l$-th order responses of the charge density, phase and Higgs modes, respectively. Then, one correspondingly finds the amplitudes of $\mu_H$ and $\mu_{\rm eff}$ as well as ${\bf p}_s$ in the $l$-th-order response as
\begin{eqnarray}
  \mu_{H}^{(l)}&=&V_q\delta{n}^{(l)},\\
  \mu_{\rm eff}^{(l)}&=&e\phi_0({\bf x})\delta_{l,1}+il\Omega\delta\theta^{(l)}/2+\partial_{x_0}\delta\theta^{(l)}/2+\mu_H^{(l)},~~~\\
  {\bf p}^{(l)}_s&=&-i{\bf q}\delta\theta^{(l)}/2+{\bm \nabla}_{\bf x}\delta\theta^{(l)}/2-e{\bf A}_0\delta_{l,1}.
\end{eqnarray}

Particularly, it is noted that based on the gauge structure in Eqs.~(\ref{gaugestructure1}) and (\ref{gaugestructure2}) of superconductors, in the $l$-th-order response, one can choose the phase-related effective electromagnetic potential  {\small $\partial_{\mu}\delta\theta^{(l)}$} to be zero. Then, the amplitude $\delta\theta^{l}$ of the $l$-th-order response of the phase mode is spatially uniform and time-independent as a background. In this situation, it has been established in the literature\cite{AK,AK2,Ba0,pm0,Am0,Ba9,Ba10,pm5,pi1,pi4,GIKE2} that the phase mode in the linear regime $\delta\theta^{(1)}$, as a scalar quantity, responds to the longitudinal electromagnetic field solely, which experiences the Coulomb screening. Consequently, the uniform linear response of the phase mode ${\delta\theta^{(1)}}/{2}=\frac{i\Omega{e{\bar \phi}_0-\omega^2_p{i{\bf
        q}{\cdot}e{\bf A}_0}/{q^2}}}{\Omega^2-v^2_pq^2}$ becomes ${\delta\theta^{(1)}}/{2}=\frac{i\Omega{e{\bar \phi}_0-\omega^2_p{i{\bf
    q}{\cdot}e{\bf A}_0}/{q^2}}}{(\Omega^2-v^2_pq^2)(1-\omega^2_p/\Omega^2)}\approx\frac{i\Omega{e{\bar \phi}_0-\omega^2_p{i{\bf
        q}{\cdot}e{\bf A}_0}/{q^2}}}{\Omega^2-\omega^2_p}$ after considering the long-range Coulomb interaction\cite{AK,Ba0,pm0,Am0,Ba9,Ba10,pm5,AK2,GIKE2}, with $v_p$ being the velocity of the phase mode and $\omega_p$ denoting the plasma frequency. The original gapless spectrum (resonance pole) is then effectively lifted up to the high-energy plasma frequency as a consequence of the Anderson-Higgs mechanism\cite{AHM}. At this case, with $\Omega\ll\omega_p$, one has $\delta\theta^{(1)}/2\approx{i{\bf q}{\cdot}e{\bf A}_0}/{q^2}$, which cancels the unphysical longitudinal vector potential in {\small ${\bf p}_s^{(1)}={\bf q}({\bf q}\cdot{e}{\bf A}_0)/q^2-e{\bf A}_0$}, and then, the superconducting momentum ${\bf p}_s$ that appears in the previous theoretical descriptions such as Ginzburg-Landau equation\cite{G1} and Meissner supercurrent\cite{G1} as well as Anderson-pump effect\cite{Am1,Am2,Am3,Am4,Am7,Am8,Am9,Am10,Am11,Am14,NL7,NL8,NL9,NL10,NL11} only involves the physical transverse vector potential. Moreover, thanks to the Coulomb screening (i.e., $\mu_H^{(1)}=-2DV_q\frac{e{\bar \phi}_0+i\Omega\delta\theta^{(1)}/2}{1+2DV_q}$ with $D$ being the density of states)\cite{AK,GIKE2}, at long-wave limit, one finds $\mu_{\rm eff}^{(1)}=e{\bf E}_{\phi}\cdot{\bf x}$, in which the original longitudinal part $e{\bar \phi}_0+i\Omega\delta\theta^{(1)}$ vanishes. Consequently, considering the spatially uniform transverse fields for the optical response (i.e., ${\bf q}\rightarrow0$ and the optical electric field ${\bf E}_0=-{\bf E}_{\phi}-i\Omega{\bf A}_0$ are spatially uniform and transverse one), the linear-order component of the light-matter interaction $H_{\rm LM}$ in Eq.~(\ref{LM}) is written as
\begin{equation}
H^{(1)}_{\rm LM}=\Big[-\frac{{\hat{\bf p}}{\cdot}{e{\bf A}_0}}{m}+(e{\bf E}_{\phi}\cdot{\bf x})\tau_3\Big]e^{i\Omega{x_0}}, \label{DELM}
\end{equation}
whereas the second-order one reads
\begin{equation}\label{PELM}
H^{(2)}_{\rm LM}=\Big(\frac{e^2A^2_0}{2m}+i\Omega\delta\theta^{(2)}+\mu_H^{(2)}\Big)e^{2i\Omega{x_0}}\tau_3.      
\end{equation}
It is noted that $H^{(1)}_{\rm LM}$ represents the drive effects of the vector and scalar potentials\cite{GIKE1}. $H^{(2)}_{\rm LM}$ denotes the pump effect, in which besides the conventional contribution $H_p$ from the vector potential as mentioned in the introduction, the second-order response of the phase mode and Hartree field also play an important role.

\section{Analytic Derivation}

In this section, for the convenience of the comparison and understanding, we first briefly introduce the results of the second-order response of the collective modes from GIKE at clean limit\cite{GIKE2}, and then, separately use the Eilenberger equation and path-integral approach to derive the second-order response of Higgs mode at clean limit.

\subsection{GIKE}
\label{sec-GIKE}

In this part, we briefly introduce the results of the second-order responses of the collective modes from GIKE at clean limit\cite{GIKE2}. Particularly, we extend our previous results in Ref.~\onlinecite{GIKE2} at low temperature up to $T_c$. Specifically, the GIKE\cite{GIKE1,GIKE2} is derived from the basic Gorkov equation of $\tau_0$-Green function {\small $G_0(x,x')=-i\langle{\hat T}\psi(x)\psi^{\dagger}(x')\rangle$} based on equal-time scheme ($t=t'$)\cite{GQ2,GQ3}, with ${\hat T}$ being the chronological ordering. To retain the gauge invariance, the gauge-invariant $\tau_0$-Green function is constructed through the Wilson line\cite{Wilson}. As a result of the gauge invariance, the complete electromagnetic effects are included\cite{GIKE1} and the charge conservation is naturally satisfied\cite{GIKE2} in the GIKE. 

In this microscopic approach,
the response of system is described by density matrix $\rho_{\bf k}=\rho^{(0)}_{\bf k}+\delta\rho_{\bf k}(R)$ in Nambu space, which consists of the equilibrium part {\small
  $\rho^{(0)}_{\bf k}=\frac{1}{2}+\frac{f(E_{\bf k}^+)-f(E_{\bf
      k}^-)}{2}(\frac{\xi_k}{E_{\bf k}}\tau_3+\frac{\Delta_{0}}{E_{\bf k}}\tau_1)$} and nonequilibrium one {\small $\delta\rho_{\bf k}(R)$}. Here, $R=(x+x')/2=(t,{\bf R})$ represents the center-of-mass coordinate; {\small $f(x)$} denotes the Fermi-distribution function; $E_{\bf k}^{\pm}$ represents the quasi-electron and quasi-hole energies, which in the presence of superconducting momentum is written as\cite{FF4,FF5,FF6,FF8,FF9,GIKE1}
\begin{equation}\label{QE}
E_{\bf k}^{\pm}={\bf v}_{\bf k}\cdot{\bf p}_s\pm{E_{\bf k}}.  
\end{equation}
Here, $E_{\bf k}=\sqrt{\xi_{\bf k}^2+\Delta_{0}^2}$ is the Bogoliubov quasiparticle energy and ${\bf v}_{\bf k}\cdot{\bf p}_s$ denotes the Doppler shift\cite{FF4,FF5,FF6,GIKE1}, with the group velocity ${\bf v}_{\bf k}=\partial_{\bf k}\xi_{\bf k}$. The nonequilibrium $\delta\rho_{\bf k}$ can be solved from the GIKE\cite{GIKE2}:
\begin{eqnarray}
&&\!\!\!\!\!\!\partial_t\rho_{\bf
    k}\!+\!i\Big[\Big(\xi_k\!+\!\frac{p_s^2}{2m}\!+\!\mu_{\rm eff}\!+\!\mu_{H}\Big)\tau_3\!+\!\Delta_0\tau_1\!+\!\delta|\Delta|\tau_1,\rho_{\bf 
k}\Big]\nonumber\\
&&\!\!\!\!\!\!\mbox{}\!-\!\frac{i}{8}\left[({\bm \nabla}_{\bf R}\!+\!2i{\bf p}_s\tau_3)({\bm
    \nabla}_{\bf R}\!\!+\!\!2i{\bf p}_s\tau_3)|\Delta|\tau_1,\partial_{\bf k}\partial_{\bf k}\rho_{\bf
k}\right]\nonumber\\
&&\!\!\!\!\!\!\mbox{}\!+\!\frac{1}{2}\left\{e{\bf E}\tau_3\!\!-\!\!({\bm \nabla}_{\bf R}\!\!+\!\!2i{\bf p}_s\tau_3)|\Delta|\tau_1,\partial_{\bf k}\rho_{\bf
k}\right\}\!-\!\Big[\frac{i\nabla^2_{\bf R}}{8m}\tau_3,\rho_{\bf 
k}\Big]\nonumber\\
&&\!\!\!\!\!\!\mbox{}\!+\!\Big\{\frac{{\bf k}\!\cdot\!{\bm \nabla}_{\bf R}}{2m}\tau_3,\rho_{\bf
    k}\Big\}\!-\!\Big[\frac{{\nabla_{\bf R}}\!\circ\!{\bf p}_s}{4m}\tau_3,\tau_3\rho_{\bf
    k}\Big]=\partial_{t}\rho_{\bf k}|_{\rm scat},
\label{GE}
\end{eqnarray}
where we have applied the unitary transformation in Eq.~(\ref{utrp}) to effectively remove the phase mode from the order parameter. Here,  {\small ${\nabla_{\bf R}}\!\circ\!{\bf p}_s=(2{\bf p}_s\!\cdot\!{\bm \nabla}_{\bf R}\!+\!{\bm
    \nabla}_{\bf R}\!\cdot\!{\bf p}_s)$}; the electric field {\small $e{\bf E}=-{\bm \nabla}_{\bf R}(e\phi+\mu_H)-\partial_te{\bf A}=-{\bm \nabla}_{\bf R}\mu_{\rm eff}+\partial_t{\bf p}_s$}.

The gauge-invariant density and current
read $n=\sum_{\bf k}(1+2\rho_{{\bf k}3})$ and ${\bf
  j}=\sum_{\bf k}\big(\frac{e{\bf k}}{m}\rho_{{\bf k}0}\big)$,
respectively. Moreover, after the unitary transformation, the equation of the order parameter becomes\cite{GIKE2}
\begin{eqnarray}
U{\sum_{\bf k}}'\rho_{{\bf k}1}&=&-|\Delta|\label{gap},\\
U{\sum_{\bf k}}'\rho_{{\bf k}2}&=&0\label{phase},
\end{eqnarray}
where $U$ denotes the pairing potential and $\rho_{{\bf k}i}$ stands for the $\tau_i$ component of $\rho_{\bf k}$; $\sum_{\bf k}'$ here and hereafter stands for the summation restricted in the spherical shell ($|\xi_{\bf k}|\le\omega_D$) with $\omega_D$ being the Debye frequency. It is noted that Eq.~(\ref{gap}) gives the gap equation and
hence the Higgs mode, whereas Eq.~(\ref{phase}) determines the phase
fluctuation as revealed in our previous work\cite{GIKE2}.

For the weak probe, by expanding $\delta\rho_{\bf k}=\delta\rho^{(1)}_{\bf
    k}+\delta\rho^{(2)}_{\bf k}$ with {\small $\delta\rho^{(1)}_{\bf k}$}
and {\small $\delta\rho^{(2)}_{\bf k}$} being the first and second order responses to
 optical probe, the GIKE becomes a chain of equations, as its first order
only involves {\small $\delta\rho^{(1)}_{\bf k}$} and equilibrium {\small
  $\rho^{(0)}_{\bf k}$} and its second order involves {\small
  $\delta\rho^{(2)}_{\bf k}$}, {\small $\delta\rho^{(1)}_{\bf k}$} and {\small
  $\rho^{(0)}_{\bf k}$}.
Consequently, starting from the lowest order, one can calculate {\small $\delta\rho^{(1)}_{\bf k}$} and {\small $\delta\rho^{(2)}_{\bf k}$} in sequence, and then, obtain the linear and second-order responses of the Higgs (phase) mode by substituting the solved {\small $\delta\rho^{(1)}_{\bf k}$} and {\small $\delta\rho^{(2)}_{\bf k}$} into Eq.~(\ref{gap}) [Eq.~(\ref{phase})], respectively.

As revealed in our previous work\cite{GIKE2}, the linear response of the Higgs mode vanishes in the long-wave limit, whereas the linear response of the phase mode recovers the previous results\cite{AK,Ba0,pm0,Am0,Ba9,Ba10,pm5} of the Anderson-Higgs mechanism\cite{AHM} mentioned in Sec.~\ref{sec-H}. Here, we present the second-order responses of the Higgs $\delta|\Delta|^{(2)}$ and phase $\delta\theta^{(2)}$ modes as well as charge-density fluctuation $\delta{n}^{(2)}$ derived from the GIKE at clean limit (the specific derivation can be found in Ref.~\onlinecite{GIKE2}):
\begin{equation}
4(\Delta_0^2-\Omega^2)\beta_g\delta|\Delta|^{(2)}\!\!=\!-\frac{2\Delta_0v_F^2}{3}[\gamma_g{e^2_{\mu_{\rm eff}}}\!+\!(p_s^{(1)})^2\lambda_g],\label{shiggs}
\end{equation}
and
\begin{equation}
  -i\Omega\delta\theta^{(2)}\!-\mu_{\rm H}^{(2)}\!=\!\frac{{p_s^2}}{2m}\!-\!\frac{1}{3mu_g}({w_g}{e^2_{\mu_{\rm eff}}}\!+t_gp_s^{(1)}e_{\mu_{\rm eff}}),\label{sphase}  
\end{equation}
as well as
\begin{equation}\label{scdf1}
\delta{n^{(2)}}=0,  
\end{equation}
with the amplitude-correlation coefficient 
\begin{equation}\label{bg1}
  \beta_g={\sum_{\bf k}}'\Big[\frac{1-2f(E_{\bf k})}{2E_{\bf k}}\frac{1}{E_{\bf k}^2-\Omega^2}+\frac{\partial_{E_{\bf k}}f(E_{\bf k})}{E^2_{\bf k}}\Big],
\end{equation}
amplitude-response coefficients
\begin{eqnarray}
 \gamma_g&=&{\sum_{\bf
    k}}'\frac{\xi_{\bf k}}{E_{\bf k}^2-\Omega^2}\partial^2_{\xi_{\bf k}}\Big[{\xi_{\bf k}}\frac{2f(E_{\bf k})-1}{2E_{\bf k}}\Big], \label{gg1}  \\
\lambda_g&=&{\sum_{\bf
    k}}'\frac{\partial^2_{E_{\bf k}}f(E_{\bf k})}{E_{\bf k}}, \label{lg1}
\end{eqnarray}
and the phase-response coefficients  
\begin{eqnarray}
u_g&=&{\sum_{\bf k}}'\frac{\Delta^2_0}{E_{\bf k}^2-\Omega^2}\frac{2f(E_{\bf k})-1}{E_{\bf k}},\label{ug}\\
t_g&=&{\sum_{\bf k}}'\frac{\Delta^2_0}{E_{\bf k}^2-\Omega^2}\partial_{\xi_{\bf k}}\Big[\xi_{\bf k}\frac{2f(E_{\bf k})-1}{E_{\bf k}}\Big],\\
w_g&=&{\sum_{\bf k}}'\frac{2\Delta^2_0}{E_{\bf k}^2-\Omega^2}(\xi_{\bf k}\partial^2_{\xi_{\bf k}}+\partial_{\xi_{\bf k}})\Big[\xi_{\bf k}\frac{2f(E_{\bf k})-1}{E_{\bf k}}\Big],~~~
\end{eqnarray}
as well as $e_{\mu_{\rm eff}}=\nabla_{\bf R}{\mu^{(1)}_{\rm eff}}/(i\Omega)$. It is noted that in our previous work\cite{GIKE2} which considers the low temperature and weak optical probe, we neglect the Doppler shift in the quasiparticle energy [Eq.~(\ref{QE})] by assuming $v_kp_s<\Delta_0$. In the present work, we sublate this approximation in order to extend the calculation to the entire temperature regime. The considered Doppler shift does not influence the previous calculation, but causes a additional contribution [i.e.,the second term on the right-hand side of Eq.~(\ref{shiggs})] in the second-order response of the Higgs mode $\delta|\Delta|^{(2)}$ through $\rho_{\bf k}^{(0)}$ in Eq.~(\ref{gap}). From Eqs.~(\ref{shiggs}) and (\ref{sphase}), one finds that the second-order responses of the Higgs $\delta|\Delta|^{(2)}$ and phase $\delta\theta^{(2)}$ modes are decoupled, as they represent mutually orthogonal excitations in the Mexican-hat potential of free energy. It is also noted that all source terms on the right-hand side of both Eqs.~(\ref{shiggs}) and (\ref{sphase}) are gauge-invariant.

{\sl Higgs-mode generation}.---Considering the spatially uniform transverse fields for the optical response and choosing the phase-related effective electromagnetic potential  {\small $\partial_{\mu}\delta\theta^{(1)}$} to be zero, Eq.~(\ref{shiggs}) following the analysis of the light-matter interaction $H_{\rm LM}$ in Sec.~\ref{sec-H} becomes
\begin{equation}
4(\Delta_0^2-\Omega^2)\beta_g\delta|\Delta|^{(2)}\!\!=\!-\frac{e^2v_F^2}{3}2\Delta_0\Big[\frac{\gamma_gE^2_{\phi}}{(i\Omega)^2}\!+\!{A_0^2}\lambda_g\Big],\label{shiggs1}
\end{equation}
It is pointed out that the right-hand side of Eq.~(\ref{shiggs1}) arises from the second order of the drive effects $H_{LM}^{(1)}$ of vector potential and scalar potential, whereas the pump effect $H_{LM}^{(2)}$ makes no contribution, in consistency with the vanishing correlation between amplitude mode and pump effect as revealed by previous symmetry analysis\cite{symmetry}.

Consequently, from Eq.~(\ref{shiggs1}), one immediately finds a finite second-order response of the Higgs mode at clean limit, contributed solely by the drive effects. Actually, according to the analysis of the Ginzburg-Landau theory as mentioned in the introduction, this finite second-order response derived from the GIKE is expected, since the GIKE near $T_c$ can recover the Ginzburg-Landau equation\cite{GIKE1}. Particularly, from Eqs.~(\ref{bg1}) and (\ref{lg1}), at low frequency, near $T_c$, one has 
\begin{eqnarray}
  \beta_g\!\!&\approx&\!\!D\int{d\xi_{\bf k}}\frac{1}{|\xi_{\bf k}|}\partial_{|\xi_{\bf k}|}\Big[\frac{2f(|\xi_{\bf k}|)-1}{2|\xi_{\bf k}|}\Big]\nonumber\\
&=&DT\sum_{\omega_n}\int{d\xi_{\bf k}}\frac{1}{|\xi_{\bf k}|}\partial_{|\xi_{\bf k}|}\left[\frac{1}{(i\omega_n)^2-\xi_{\bf k}^2}\right]\nonumber\\
&=&DT\sum_{\omega_n}\int{d\xi_{\bf k}}\frac{2}{[(\omega_n)^2+\xi_{\bf k}^2]^2}=\frac{7DR(3)}{4(\pi{T})^2},~~~ \label{bg} \\
\lambda_g\!\!&\approx&\!\!D\int{d\xi_{\bf k}}\frac{\partial_{|\xi_{\bf k}|}^2f(|\xi_{\bf k}|)}{|\xi_{\bf k}|}\!=\!\sum_{\omega_n}\int{d\xi_{\bf k}}\frac{2DT}{(i\omega_n\!-\!|\xi_{\bf k}|)^3|\xi_{\bf k}|}\nonumber\\
&=&4DT\sum_{\omega_n}\int{d\xi_{\bf k}}\frac{3\omega_n^2-\xi_{\bf k}^2}{(\omega^2_n+\xi^2_{\bf k})^3}=\frac{7DR(3)}{2(\pi{T})^2},~~~ \label{lg}
\end{eqnarray}
with $\omega_n=(2n+1)\pi{T}$ being the Matsubara frequency. Then, considering the vector potential alone, Eq.~(\ref{shiggs1}) derived from the GIKE exactly recovers Eq.~(\ref{GL-HME}) derived from the Ginzburg-Landau theory.

Moreover, it is noted in Eq.~(\ref{shiggs1}) that the drive effect $E_{\phi}$ of the scalar potential also contributes to the Higgs-mode generation. This contribution, being finite from $T=0$ to $T=T_c$, is different from the one of the vector potential that emerges only at finite temperature\cite{DS1}. Actually, this difference is natural, since the superconductors can directly respond to vector potential ${\bf A}$ (Meissner effect/Ginzburg-Landau kinetic term) in addition to the electric field ${\bf E}=-{\nabla}_{\bf R}\phi-\partial_t{\bf A}$ (optical-electric-field effect), differing from normal metals that solely respond to electric field. Consequently, the contribution of the scalar potential captures the optical-electric-field effect, whereas the contribution of the vector potential characterizes the electromagnetic effects including the Meissner effect/Ginzburg-Landau kinetic term as well as the optical-electric-field effect. As mentioned in Sec.~\ref{sec-H}, according to the gauge structure [Eqs.~(\ref{gaugestructure1}) and (\ref{gaugestructure2})] in superconductors, among the scalar potential, vector potential as well as the phase-related effective electromagnetic potential  {\small $\partial_{\mu}\delta\theta$}, one can only choose one quantity to be zero in superconductors. Consequently, the inclusion of the contribution from the scalar potential here is essential, since we have chosen zero {\small $\partial_{\mu}\delta\theta^{(1)}$}.

{\sl Phase-mode generation}.---For spatially uniform transverse optical fields and zero {\small $\partial_{\mu}\delta\theta^{(1)}$}, Eq.~(\ref{sphase}) becomes
\begin{equation}
  -i\Omega\delta\theta^{(2)}\!-\mu_{\rm H}^{(2)}\!=\!\frac{{e^2A_0^2}}{2m}\!-\!\frac{e^2}{3m}\Big[\frac{w_g}{u_g}\frac{E^2_{\phi}}{(i\Omega)^2}\!+\!\frac{t_g}{u_g}\frac{E_{\phi}A_0}{i\Omega}\Big].\label{sphase1}  
\end{equation}
The first term on the right-hand side of Eq.~(\ref{sphase1}) arises from the pump effect $H_p$ of vector potential and the last two terms come from the drive effect $H_{\rm LM}^{(1)}$. The finite contribution of the pump effect here agrees with the finite correlation between phase mode and $H_p$ as revealed by previous symmetry analysis\cite{symmetry}. 

Consequently, from Eqs.~(\ref{scdf1}) and (\ref{sphase1}), one finds in the second-order optical response a vanishing charge-density fluctuation $\delta{n^{(2)}}$ but a finite phase-mode generation $\delta\theta^{(2)}$, respectively. As mentioned in the introduction, the vanishing charge-density fluctuation $\delta{n^{(2)}}$ agrees with the inversion symmetry and charge conservation. As for the phase-mode generation, the Hartree field on the left-hand side of Eq.~(\ref{sphase1}) vanishes as $\mu_H^{(2)}=2V_q\delta{n}^{(2)}=0$, whereas the right-hand side of the equation is determined by the transverse optical field and hence free from the influence of the Coulomb screening. As we pointed out in Ref.~\onlinecite{GIKE2}, this phase-mode generation, showing a spatially uniform but temporally oscillating phase, is a unique feature of the optical properties in the second-order response, and does not manifest itself or incur any consequence in the thermodynamic, electric or magnetic properties. Nevertheless, we show in the following Sec.~\ref{sec-PIA} that this phase-mode generation $\delta\theta^{(2)}$ that has long been overlooked in the literature is essential in the theoretical description of the second-order optical response, since $\delta\theta^{(2)}$ provides an effective field to exactly cancel the unphysical excitation of the charge-density fluctuation  reported by Cea {\em et al.}\cite{Cea1,Cea2,Cea3} and hence guarantee the charge conservation.

It is stressed that all results from the GIKE, including the Higgs-mode generation [Eq.~(\ref{shiggs1})] from both contributions of the drive effects of scalar and vector potentials and phase mode generation [Eq.~(\ref{sphase1})] as well as vanishing charge-density fluctuation [Eq.~(\ref{scdf1})] can be exactly recovered from the path-integral approach within the gauge-invariant manner, to be shown in the following Sec.~\ref{sec-PIA}.

\subsection{Eilenberger equation}
\label{sec-Eie}

Following the previous work by Silaev\cite{Silaev}, we next use the Eilenberger equation\cite{Eilen,Ba20,Eilen1} to derive the second-order optical response of the Higgs mode. The Eilenberger equation\cite{Eilen,Ba20,Eilen1} is derived from the basic Gorkov equation of $\tau_3$-Green function {\small $G_3(x,x')=-i\tau_3\langle{\hat T}\psi(x)\psi^{\dagger}(x')\rangle$} through the quasiclassical approximation\cite{QA1}:
\begin{equation}\label{QG1}
g(x_0,x_0',{\bf R},{{\bf k}}_F)=\frac{i}{\pi}\int{d\xi_{\bf k}}[G_3(x_0,x_0',{\bf R},{{\bf k}})].  
\end{equation}
Here, $G(x_0,x_0',{\bf R},{{\bf k}})=\int{d{\bf r}}G(x,x')e^{-i{\bf k}\cdot({\bf x}-{\bf x'})}$. In the imaginary time domain ($x_0\rightarrow{i\tau_1}$, $x_0'\rightarrow{i\tau_2}$), in consideration of the spatially uniform transverse vector potential alone, the Eilenberger equation at clean limit reads\cite{Silaev}:
\begin{equation}\label{ELE}
\{\tau_3\partial_{\tau},{\hat g}\}_{\tau}-[(\Delta_0+\delta|\Delta|)\tau_1\tau_3,{\hat g}]_{\tau}+[e{\bf A}\cdot{\bf v}_F\tau_3,{\hat g}]_{\tau}\!=\!0,  
\end{equation}
where {\small $[X,{\hat g}]_{\tau}=X(\tau_1)g(\tau_1,\tau_2)-g(\tau_1,\tau_2)X(\tau_2)$} and {\small $\{X,{\hat g}\}_{\tau}=X(\tau_1)g(\tau_1,\tau_2)+g(\tau_1,\tau_2)X(\tau_2)$}. Moreover, the Eilenberger equation is supplemented by the normalization condition\cite{Silaev,Ba20,Eilen1} $\int{d\tau}g(\tau_1,\tau)g(\tau,\tau_2)=1$. While the corresponding gap equation is written as\cite{Eilen,Silaev,Ba20,Eilen1} 
\begin{equation}\label{ELGE}
\Delta=U{\rm Tr}[\langle{g(\tau,\tau)}\rangle_F\tau_2/2],  
\end{equation}
with $\langle...\rangle_F$ denoting the angular average over the Fermi surface.

By self-consistently solving Eqs.~(\ref{ELE})-(\ref{ELGE}), one can formulate the Higgs-mode generation at clean limit. Specifically, in the optical response with ${\bf A}(\tau)={\bf A}_0e^{i\Omega\tau}$, the quasiclassical Green function is given by ${\hat g}={\hat g}^{(0)}+{\hat g}^{(1)}+{\hat g}^{(2)}$ with the $m$-th order response ${\hat g}^{(m)}$ written as
\begin{equation}
{\hat g}^{(m)}\!\!=\!T\!\sum_{\omega_n}g^{(m)}(i\omega_n\!+\!im\Omega,i\omega_n)e^{i(\omega_n+m\Omega)\tau_1-i\omega_n\tau_2}.
\end{equation}
Consequently, the Eilenberger equation in Eq.~(\ref{ELE}) becomes a chain of equations, whose first order only involves $g^{(0)}$ and $g^{(1)}$ and second order involves $g^{(0)}$ and  $g^{(1)}$ as well as $g^{(2)}$. Then, with the equilibrium $g^{(0)}(i\omega_n)=\frac{\omega_n\tau_3-\Delta_0\tau_2}{\sqrt{\omega_n^2+\Delta_0^2}}$, one can solve $g^{(1)}$ and $g^{(2)}$ in sequence, whose specific expressions are given by (the detailed derivation can be found in Ref.~\onlinecite{Silaev})
\begin{eqnarray}
 &&\!\!\!\!\!\!g^{(1)}(i\omega_{n}\!+\!i\Omega,i\omega_n)=\!-\delta\Delta^{(1)}\Gamma_2(i\omega_{n}\!+\!i\Omega,i\omega_n)\nonumber\\
  &&\!\!\!\!\!\!\mbox{}+\!i(e{\bf A}_0\!\cdot\!{\bf v}_F)\Gamma_3(i\omega_{n}\!+\!i\Omega,i\omega_n), \label{SELR1}\\
 &&\!\!\!\!\!\!g^{(2)}(i\omega_{n}\!\!+\!2i\Omega,i\omega_n)=-\delta\Delta^{(2)}\Gamma_2(i\omega_{n}\!+\!2i\Omega,i\omega_n)\!\!\nonumber\\
&&\!\!\!\!\!\!\mbox{}+\!(e{\bf A}_0\!\cdot\!{\bf v}_F)^2\frac{[(\omega_n\!\!+\!2\Omega)\tau_3\!+\!\Delta_0\tau_2]{\hat \Xi}\!+\!{\hat \Xi}(\omega_n\tau_3\!+\!\Delta_0\tau_2)}{(\omega_n\!+\!2\Omega)^2\!-\!\omega^2_n},~~~~~\label{SELR2}  
\end{eqnarray}
with the correlation function 
\begin{equation}
  \Gamma_j(i\omega_n\!+\!i\Omega,i\omega_n)=\frac{g^{(0)}(i\omega_n+i\Omega)\tau_jg^{(0)}(i\omega_n)\!-\!\tau_j}{\sqrt{(\omega_n+\Omega)^2\!+\!\Delta_0^2}\!+\!\sqrt{\omega^2_n\!+\!\Delta_0^2}},
\end{equation}
and {\small ${\hat \Xi}=\Gamma_3(i\omega_{n}\!+\!2i\Omega,i\omega_n\!+\!i\Omega)\tau_3\!-\!\tau_3\Gamma_3(i\omega_{n}\!+\!i\Omega,i\omega_n)$}.\\

{\sl Higgs-mode generation}.---Substituting the solved $g^{(1)}$ and $g^{(2)}$ into the gap equation [Eq.~(\ref{ELGE})], one can obtain the linear and second-order response of the Higgs mode, respectively.  The linear response $\delta\Delta^{(1)}=0$, as the anisotropic source term from the vector potential on the right-hand side of Eq.~(\ref{SELR1}) vanishes after the angular average over the Fermi surface. The obtained second-order response of the Higgs mode is written as
\begin{equation}
(4\Delta_0^2\!+\!4\Omega^2)\delta|\Delta|^{(2)}\beta_E=-\frac{e^2A_0^2v_F^22\Delta_0}{3}\lambda_E,  \label{Ei-HME}
\end{equation}
with the amplitude-correlation coefficient $\beta_E$ and the amplitude-response coefficient $\lambda_E$ given by 
\begin{widetext}
\begin{eqnarray}
\beta_E&=&2\Big\{\frac{{U^{-1}\!+\!T\sum_{\omega_n}{\rm Tr}[\Gamma_2(i\omega_{n}\!+\!i2\Omega,i\omega_n)\tau_2/2]}}{(4\Delta_0^2\!+\!4\Omega^2)}\Big\}=T\sum_{\omega_n}\frac{1/[\sqrt{(\omega_n\!+\!2\Omega)^2\!+\!\Delta_0^2}\sqrt{\omega^2_n\!+\!\Delta_0^2}]}{\sqrt{(\omega_n\!+\!2\Omega)^2\!+\!\Delta_0^2}\!+\!\sqrt{\omega^2_n\!+\!\Delta_0^2}},
\end{eqnarray}  
\begin{equation}
  \lambda_E\!=\!-T\sum_{\omega_n}\frac{{\rm Tr}[i(\omega_n\!\!+\!2\Omega)\tau_1{\hat \Xi}\!-\!i\omega_n{\hat \Xi}\tau_1\!+\!2\Delta_0{\hat \Xi}]}{\Delta_0[(\omega_n\!+\!2\Omega)^2\!-\!\omega^2_n]}\!=\!-\frac{T}{2\Omega^2}\sum_{\omega_n}\Big[\frac{2}{\sqrt{(\omega_n\!+\!\Omega)^2\!+\!\Delta_0^2}}\!-\!\frac{1}{\sqrt{(\omega_n\!+\!2\Omega)^2\!+\!\Delta_0^2}}\!-\!\frac{1}{\sqrt{\omega_n^2\!+\!\Delta_0^2}}\Big].  \label{DE}
\end{equation}
\end{widetext}

It is noted that Eq.~(\ref{Ei-HME}) is exactly same as the one obtained in the previous work by Silaev\cite{Silaev}. Nevertheless, in Ref.~\onlinecite{Silaev}, the amplitude-response coefficient $\lambda_E$ is directly considered to disappear after the summation over the Matsubara frequency, leading to a zero Higgs-mode generation. However, in contrast to the discrete $\omega_n/(\pi{T})=2n+1$, the optical frequency $\Omega$ must be continuous in this circumstance (refer to Sec.~\ref{sec-discuss}). Considering this point, the amplitude-response coefficient $\lambda_E$ does not vanish. In fact, using the fact:
\begin{equation}\label{fact}
\frac{1}{\sqrt{\Delta_0^2\!+\!(\omega_n\!+\!m\Omega)^2}}=\!\!\int\frac{d\xi_{\bf k}}{\pi}\frac{1}{E_{\bf k}^2\!-\!(i\omega_n\!+\!im\Omega)^2}
\end{equation}
and considering low-frequency regime ($\Omega<E_{\bf k}$), after the standard Matsubara-frequency summations, one can find a nonzero amplitude-response coefficient
\begin{eqnarray}
 &&\!\!\!\!\!\!\lambda_E=-\frac{1}{2\Omega^2}\frac{1}{\pi}\int{d\xi_{\bf k}}\sum_{\eta=\pm}\Big[\frac{f({\eta}E_{\bf k})}{2{\eta}E_{\bf k}}+\frac{f({\eta}E_{\bf k}-2i\Omega)}{2{\eta}E_{\bf k}}\nonumber\\
    &&\!\!\!\!\!\!\mbox{}-\frac{2f({\eta}E_{\bf k}\!-\!i\Omega)}{2{\eta}E_{\bf k}}\Big]\!\approx\!\frac{1}{\pi}\!\!\int\!\!{d\xi_{\bf k}}\Big[\frac{\partial_{E_{\bf k}}^2f(E_{\bf k})}{2E_{\bf k}}\!+\!O\Big(\frac{\Omega^3}{E^3_{\bf k}}\Big)\Big].~~~~\label{le}
\end{eqnarray}
Consequently, a finite second-order response of the Higgs mode, contributed by the drive effect of the vector potential, is achieved at clean limit by using Eilenberger equation, in contrast to the previous work by Silaev\cite{Silaev}. Actually, according to the analysis of the Ginzburg-Landau theory as mentioned in the introduction, this finite second-order response of the Higgs mode derived from the Eilenberger equation is expected,  since the Eilenberger equation near $T_c$ can recover the Ginzburg-Landau equation\cite{Ba20}. Particularly, at low frequency, the amplitude-correlation coefficient 
\begin{eqnarray}
\beta_E&\approx&\!\frac{T}{2}\!\sum_{\omega_n}\frac{1}{(\Delta_0^2+\omega^2_n)^{3/2}}\!=\!-\frac{\partial_{\Delta_0}}{2\Delta_0}\Big[T\!\sum_{\omega_n}\frac{1}{\sqrt{\Delta_0^2+\omega^2_n}}\Big] \nonumber\\
&=&\!2\int\frac{d\xi_{\bf k}}{\pi}\frac{\partial_{E_{\bf k}}}{4E_{\bf k}}\Big[\frac{2f(E_{\bf k})\!-\!1}{2E_{\bf k}}\Big].  \label{be}
\end{eqnarray}
Then, by comparing Eqs.~(\ref{bg})-(\ref{lg}) and Eqs.~(\ref{be})-(\ref{le}), one has $\beta_E=\beta_g/(2D\pi)$ and $\lambda_E=\lambda_g/(2D\pi)$. Therefore, for real optical frequency ($i\Omega\rightarrow\Omega$), Eq.~(\ref{Ei-HME}) derived from the Eilenberger equation exactly recovers Eq.~(\ref{shiggs1}) derived from the GIKE, and hence, near $T_c$, can also recover Eq.~(\ref{GL-HME}) derived from the Ginzburg-Landau theory.

Nevertheless, it is noted from Eq.~(\ref{ELE}) that the quasiclassical Eilenberger equation\cite{Eilen,Ba20,Eilen1}  only involves the drive effect of the vector potential, i.e., the first term of $H^{(1)}_{\rm LM}$ in Eq.~(\ref{DELM}). Whereas the drive effect of the scalar potential, i.e., the second term of $H^{(1)}_{\rm LM}$ in Eq.~(\ref{DELM}), is hard to handle in the quasiclassical formalism due to its spatial dependence (${\hat {\bf x}}\rightarrow{-i{\hat \partial}_{\bf k}}$). Hence, the finite contribution from the drive effect of the scalar potential to the Higgs-mode generation, which is nonzero at $T=0$, is generically dropped out in this approach. Furthermore,  even in consideration of the vector potential alone, the density-related pump effect $H_p$ is generically dropped out in the quasiclassical Eilenberger equation. Accordingly, the response of the density-related phase mode, i.e., the second term of $H^{(2)}_{\rm LM}$ in Eq.~(\ref{PELM}), as well as the Hartree field that is related to the charge-density fluctuation and long-range Coulomb interaction, are also dropped out. In fact, these deficiencies are because that the Eilenberger equation in Eq.~(\ref{ELE}) is not gauge-invariant, and hence, the contained electromagnetic effect is incomplete. 

\subsection{Path-integral approach}
\label{sec-PIA}

Following the previous work by Cea {\em et al.}\cite{Cea1,Cea2,Cea3}, we next use the path-integral approach to derive the second-order optical response of the Higgs mode. We start with the generalized action of superconductors in the presence of electromagnetic potential $A_{\mu}$ \cite{Ba0,G1}:
\begin{eqnarray}
&&\!\!\!\!\!\!S[\psi,\psi^*]=\int{dx}\Big[\sum_{s=\uparrow,\downarrow}\psi^*_s(x)(i\partial_t-\xi_{\hat {\bf p}-e{\bf A}}-e\phi)\psi_s(x)\nonumber\\
   &&\!\!\!\!\!\!\mbox{}+\!U\psi^*_{\uparrow}(x)\psi^*_{\downarrow}(x)\psi_{\downarrow}(x)\psi_{\uparrow}(x)\!-\!\frac{1}{2}\!\!\int\!{dx'}V(x\!-\!x')n(x)n(x')\Big],\nonumber\\
\end{eqnarray}
differing from the one used in Refs.~\onlinecite{Cea1,Cea2,Cea3} with the vector potential alone. Here, the density $n(x)=\sum_{s=\uparrow,\downarrow}\psi^{*}_s(x)\psi_s(x)$. After the Hubbard-Stratonovich transformation, one has
\begin{eqnarray}
&&\!\!\!\!\!S[\psi,\psi^*]=\!\!\int{dx}\bigg[\sum_{s=\uparrow,\downarrow}\!\!\psi^*_s(x)(i\partial_t\!-\!\xi_{\hat {\bf p}-e{\bf A}}\!-\!e\phi\!-\!\mu_H)\psi_s(x)\nonumber\\
&&\mbox{}+\!\psi^{\dagger}(x){\hat \Delta}(x)\psi(x)\!-\!\frac{|\Delta(x)|^2}{U}\bigg]\!+\!\frac{1}{2}\sum_{Q}\frac{|\mu_H(Q)|^2}{V_Q}.~~~~~~\label{sis}
\end{eqnarray}
Here, $\mu_H$ stands for the auxiliary field, i.e., the Hartree field that reflects the density fluctuation; $V_Q$ denotes the Fourier component of $V(x-x')$. The action in Eq.~(\ref{sis}) satisfies the gauge structure in Eqs.~(\ref{gaugestructure1})-(\ref{gaugestructure2}) revealed by Nambu\cite{gi0} and hence is gauge invariant. 

By further using the unitary transformation in Eq.~(\ref{utrp}) to effectively remove the phase mode from the order parameter, the action in Eq.~(\ref{sis}) becomes
\begin{eqnarray}
S[\psi,\psi^*]&=&\int{dx}\Big\{\psi^*(x)[G_0^{-1}(x)\!-\!\Sigma(x)]\psi(x)\!-\!\frac{|\Delta(x)|^2}{U}\nonumber\\
  &&\mbox{}-\eta_f{\rm Tr}[\Sigma(x)\tau_3/2]\Big\}+\frac{1}{2}\sum_{Q}\frac{|\mu_H(Q)|^2}{V_Q}.
\end{eqnarray}
where $\eta_f=\sum_{\bf k}1$ emerges due to the anti-commutation of the Fermi field; the Green function $G_0^{-1}(x)=i\partial_t-H_0$, which in frequency-momentum space [$x\rightarrow{p=(p_0,{\bf k})}$] reads $G_0(p)=(p_0+\xi_{\bf k}\tau_3+\Delta_0\tau_1)/(p_0^2-E_{\bf k}^2)$ and the self-energy $\Sigma(x)=H_{\rm LM}+\delta|\Delta|\tau_1+\mu_{H}\tau_3$.  

After the standard integration over the Fermi field, one has $S=S_0+{S}_{\rm ne}[A_{\mu},\delta|\Delta|,\delta\theta]$, consisting of the equilibrium part $S_0$ and the non-equilibrium one:
\begin{eqnarray}
&&\!\!\!\!\!S_{\rm ne}[A_{\mu},\delta|\Delta|,\delta\theta]\!=\!-\sum_{n=1}^{\infty}\frac{1}{n}{\rm {\bar Tr}}[(G_0\Sigma)^n]\!+\!\frac{1}{2}\sum_{Q}\frac{|\mu_H(Q)|^2}{V_Q}\nonumber\\
 &&\!\!\!\!\mbox{}-\!\!\int{dx}\eta_f{\rm Tr}[\Sigma(x)\tau_3/2]-\!\int{dx}\frac{(\delta|\Delta|)^2+2\Delta_0\delta|\Delta|}{U}. 
\end{eqnarray}
In the non-equilibrium $S_{\rm ne}[A_{\mu},\delta|\Delta|,\delta\theta]$, the anisotropic linear and third orders with respect to the electromagnetic potential vanish after the angular integration of momentum, whereas the second-order part that corresponds to the linear current excitation has been well established in the literature\cite{pi1,pi4}. To discuss the experimentally observed third-harmonic current, one needs to formulate the expansion of the action with respect to the fourth order of the electromagnetic potential, and hence, keeps the expansions up to $n=4$. For the convenience of the derivation, we consider the spatially uniform transverse optical fields and choose zero phase-related effective electromagnetic potential  {\small $\partial_{\mu}\delta\theta^{(l)}$}. Then, following the analysis of the light-matter interaction $H_{\rm LM}$ in Sec.~\ref{sec-H}, the related action with expansions up to $n=4$ is written as
\begin{widetext}
\begin{eqnarray}
  S\big[A^4_{\mu}\big]&=&-\frac{1}{2}{\rm {\bar Tr}}[G_0(H^{(2)}_{\rm LM}\!+\!\delta|\Delta|\tau_1\!+\!\mu_{H}\tau_3)G_0(H^{(2)}_{\rm LM}\!+\!\delta|\Delta|\tau_1\!+\!\mu_{\rm H}\tau_3)]\!-\!{\rm {\bar Tr}}[G_0H_{\rm LM}^{(1)}G_0H_{\rm LM}^{(1)}G_0(H^{(2)}_{\rm LM}\!+\!\delta|\Delta|\tau_1\!+\!\mu_{\rm H}\tau_3)]\nonumber\\
  &&\mbox{}\!-\!\frac{1}{4}{\rm {\bar Tr}}[G_0H_{\rm LM}^{(1)}G_0H_{\rm LM}^{(1)}G_0H_{\rm LM}^{(1)}G_0H_{\rm LM}^{(1)}]\!-\!\sum_{Q}\frac{|\delta|\Delta|_Q|^2}{U}\!+\!\frac{1}{2}\sum_{Q}\frac{|\mu_H(Q)|^2}{V_Q}\nonumber\\
  &=&-\!\sum_{Q}\Big[\chi_H|\delta|\Delta|_{2Q}|^2\!+\!\chi_{33}\Big|\frac{e^2A_0^2}{2m}\!+\!\mu_H\!+\!iQ_0\delta\theta^{(2)}\Big|^2\!-\!\frac{|\mu_H|^2}{2V_Q}\Big]\!-\!\sum_{Q}\Bigg\{\!2\chi_{13}\Big(\frac{e^2A_0^2}{2m}\!+\!\mu_H\!+\!iQ_0\delta\theta^{(2)}\Big)_{2Q}\delta|\Delta|_{-2Q}  \nonumber\\
  &&\mbox{}\!+\!\Big[\chi_{001}\frac{e^2A^2_0v_F^2}{3}\!+\!(\chi_{0{\bar 3}1}-\chi_{3{\bar 0}1})\frac{eA_0eE_{\phi}v_F^2}{3i}+\chi_{3{\bar{\bar 3}}1}\frac{e^2E_{\phi}^2v_F^2}{3}\Big]_{2Q}\delta\Delta_{-2Q}\!+\!\Big(\frac{e^2A_0^2}{2m}\!+\!\mu_H\!+\!iQ_{0}\delta\theta^{(2)}\Big)_{-2Q}\nonumber\\
  &&\mbox{}\!\times\!\Big[\chi_{003}\frac{e^2A^2_0v_F^2}{3}\!+\!(\chi_{0{\bar 3}3}-\chi_{3{\bar 0}3})\frac{eA_0eE_{\phi}v_F^2}{3i}+\chi_{3{\bar{\bar 3}}3}\frac{e^2E_{\phi}^2v_F^2}{3}\!+\!(\chi_{0{3}3}-\chi_{303})\frac{eA_0eE_{\phi}}{3mi}+\chi_{3{\bar 3}3}\frac{e^2E_{\phi}^2}{3m}\Big]_{-2Q}\!+\!h.c.\Bigg\}\nonumber\\
  &&\mbox{}\!-\!\frac{e^4v_F^4}{20}{\rm {\bar Tr}}\{[G_0(A_0-E_{\phi}\tau_3\partial_{\xi_{\bf k}})]^4\}, \label{sis2}
\end{eqnarray}
\end{widetext}
in which we have considered a large $v_F$ (i.e., neglected the terms proportional to $v_F^2$ and $v_F^0$ and only kept the ones proportional to $v_F^4$) in the expansion of $n=4$. Here, the frequency-momentum vector $Q=(Q_0,{\bf Q})$; $\chi_H=\chi_{11}+1/U$ denotes the energy-spectrum function of the Higgs mode; the correlation coefficients are written as 
\begin{eqnarray}
\chi_{ij}\!\!&=&\!\!\!\frac{1}{2}\sum_p{\rm Tr}[G_0(p+2Q)\tau_iG_0(p)\tau_j], \label{chiij} \\
  \chi_{ijk}\!\!&=&\!\!\!\sum_p{\rm Tr}[G_0(p+2Q)\tau_iG_0(p+Q)\tau_jG_0(p)\tau_k],\\
  \chi_{i{\bar j}k}\!\!&=&\!\!\!\sum_p{\rm Tr}[G_0(p+2Q)\tau_i\partial_{\xi_{\bf k}}G_0(p+Q)\tau_jG_0(p)\tau_k],  \\
  \chi_{i{\bar{\bar 3}}k}\!\!&=&\!\!\!\frac{1}{3}\sum_p{\rm Tr}[G_0(p\!+\!2Q)\tau_i{\partial^2_{\xi_{\bf k}}}G_0(p\!+\!Q)\tau_3G_0(p)\tau_k].~~~~~~\label{chii3k}
\end{eqnarray}
It is noted that the action in Eq.~(\ref{sis2}) exactly recovers the one in the previous work by Cea {\em et al.}\cite{Cea1,Cea2,Cea3}, if one only keeps the second-order perturbation (i.e., $n=2$) expansion and neglects the third- and forth-order perturbation (i.e., $n=3$ and $n=4$) expansions. As revealed in Refs.~\onlinecite{Cea1,Cea2,Cea3}, the second-order correlations $\chi_{11}$ and $\chi_{33}$ characterize the amplitude-amplitude and density-density correlations, respectively. The density-amplitude correlation $\chi_{13}$ is zero as a consequence of the particle-hole symmetry, and hence, the only coupling between the Higgs mode and second-order optical field   in the second-order perturbation expansion, i.e., the coupling between the Higgs mode and pump effect, vanishes.

As for the third-order correlations, one can prove that $\chi_{003}$, $\chi_{0{\bar 3}3}$, $\chi_{3{\bar 0}3}$ and $\chi_{3{\bar{\bar 3}}3}$ vanish as a consequence of the particle-hole symmetry, and $\chi_{0{\bar 3}1}-\chi_{3{\bar 0}1}=0$ (refer to Appendix~\ref{ADCC}). Then, the action in Eq.~(\ref{sis2}) is simplified as
\begin{widetext}
\begin{eqnarray}
  S\big[A^4_{\mu}\big]&=&-\!\sum_{Q}\Big\{\chi_H|\delta|\Delta|_{2Q}|^2\!+\!\Big[\Big(\chi_{001}\frac{e^2A^2_0v_F^2}{3}\!+\!\chi_{3{\bar{\bar 3}}1}\frac{e^2E_{\phi}^2v_F^2}{3}\Big)_{2Q}\delta\Delta_{-2Q}+h.c.\Big]\Big\} \nonumber\\
&&\mbox{}\!-\!\sum_Q\Big(\chi_{33}\Big|\frac{e^2A_0^2}{2m}\!+\!\mu_H\!+\!iQ_0\delta\theta^{(2)}\!-\!\frac{\zeta_1eA_0eE_{\phi}}{3miQ_0}\!+\!\frac{\zeta_2e^2E_{\phi}^2}{3mQ_0^2}\Big|^2\!+\!\frac{|\mu_H|^2}{2V_Q}\Big)\!-\!\frac{e^4v_F^4}{20}{\rm {\bar Tr}}\{[G_0(A_0-E_{\phi}\tau_3\partial_{\xi_{\bf k}})]^4\},~~~\label{Fac}
\end{eqnarray}
\end{widetext}
with $\zeta_1=Q_0\frac{\chi_{0{3}3}-\chi_{303}}{\chi_{33}}$ and $\zeta_2=Q_0^2\frac{\chi_{3{\bar 3}3}}{\chi_{33}}$. In Eq.~(\ref{Fac}), the first term is related to the Higgs mode, and the second one is related to the phase mode and charge-density fluctuation. Whereas the third one denotes the fourth order of the drive effect, which is related to the thermal effect.  

\subsubsection{Higgs-mode generation}

By using the action in Eq.~(\ref{Fac}), we discuss the Higgs-mode generation in the second-order optical response and its contribution to the third-harmonic current. Considering the optical response $A_{\mu}(Q)\rightarrow{A_{\mu}}\delta(Q_0-\Omega)\delta({\bf q})$, one has $\chi_{H}(2Q)=(4\Delta_0^2-4\Omega^2)\beta_p$ and $\chi_{001}=2\Delta_0\lambda_p$ as well as $\chi_{3{\bar{\bar 3}}1}=\frac{2\Delta_0}{(i\Omega)^2}\gamma_p$, with the coefficients $\beta_p$ and $\lambda_p$ as well as $\gamma_p$ at low frequency given by (refer to Appendix~\ref{ADCC})
\begin{eqnarray}
\beta_p&\approx&D\int{d\xi_{\bf k}}\frac{\partial_{E_{\bf k}}}{2E_{\bf k}}\Big[\frac{2f(E_{\bf k})\!-\!1}{2E_{\bf k}}\Big],\label{bp}\\
\lambda_p&=&\frac{1}{2Q_0^2}\sum_{\bf k}\Big\{\Big[\frac{f(E_{\bf k}\!-\!2Q_0)}{2E_{\bf k}}\!+\!\frac{f(E_{\bf k})}{2E_{\bf k}}\!-\!\frac{2f(E_{\bf k}\!-\!Q_0)}{2E_{\bf k}}\Big]\nonumber\\
&&\mbox{}+\{E_{\bf k}\rightarrow-E_{\bf k}\}\Big\},\label{lp}\\
\gamma_p&\approx&D\int{d\xi_{\bf k}}\frac{\xi_{\bf k}^4}{E_{\bf k}^5}\partial_{E_{\bf k}}\Big[\frac{2f(E_{\bf k})-1}{2E_{\bf k}}\Big].\label{gp}
\end{eqnarray}
Then, from Eq.~(\ref{Fac}), the equation of motion of the Higgs mode reads
\begin{equation}\label{HMEQP}
4(\Delta_0^2\!-\!\Omega^2)\beta_p\delta|\Delta|^{(2)}=-2\Delta_0\frac{e^2v_F^2}{3}\Big[\lambda_p{A^2_0}\!+\!\frac{\gamma_pE^2_{\phi}}{(i\Omega)^2}\Big].
\end{equation}
Consequently, one immediately finds a finite second-order response of the Higgs mode from the path-integral approach at clean limit, contributed by both drive effects of the vector and scalar potentials. Particularly, at low frequency, by comparing Eqs.~(\ref{bg})-(\ref{lg}) and Eqs.~(\ref{bp})-(\ref{lp}), one finds $\beta_p=\beta_g/2$ and
\begin{equation}
\lambda_p~{\approx}~D\int{d\xi_{\bf k}}\frac{\partial_{E_{\bf k}}^2f(E_{\bf k})}{2E_{\bf k}}=\lambda_g/2.   
\end{equation}
In addition, from Eq.~(\ref{gg1}) at low frequency, one has
\begin{eqnarray}
  \gamma_g&\approx&{D}\int{d\xi_{\bf k}}\frac{\xi_{\bf k}}{E_{\bf k}^2}\partial^2_{\xi_{\bf k}}\Big[{\xi_{\bf k}}\frac{2f(E_{\bf k})-1}{2E_{\bf k}}\Big]\nonumber\\
  &=&D\int{d\xi_{\bf k}}\Big(\frac{3\xi_{\bf k}^2}{E_{\bf k}^3}\!-\!\frac{\xi_{\bf k}^4}{E_{\bf k}^5}\!+\!\frac{\xi_{\bf k}^3}{E_{\bf k}^3}\partial_{\xi_{\bf k}}\Big)\partial_{E_{\bf k}}\Big[\frac{2f(E_{\bf k})\!-\!1}{2E_{\bf k}}\Big] \nonumber\\
  &=&D\int{d\xi_{\bf k}}\Big[\frac{3\xi_{\bf k}^2}{E_{\bf k}^3}\!-\!\frac{\xi_{\bf k}^4}{E_{\bf k}^5}\!-\!\partial_{\xi_{\bf k}}\Big(\frac{\xi_{\bf k}^3}{E_{\bf k}^3}\Big)\Big]\partial_{E_{\bf k}}\Big[\frac{2f(E_{\bf k})\!-\!1}{2E_{\bf k}}\Big]\nonumber\\
  &=&2\gamma_p.
\end{eqnarray}
Consequently, Eq.~(\ref{HMEQP}) derived from the path-integral approach exactly recovers Eq.~(\ref{shiggs1}) derived from the GIKE, {\sl with both contributions of the vector and scalar potentials incorporated}.

The obtained finite second-order response of the Higgs mode through the vector-potential drive effect within the path-integral approach is quite natural, since this approach near $T_c$ can also recover the Ginzburg-Landau equation\cite{PI2GL}. Near $T_c$,  Eq.~(\ref{HMEQP}) with the vector potential alone recovers Eq.~(\ref{GL-HME}) derived from the Ginzburg-Landau theory. Nevertheless, in the previous works by Cea {\em et al.}\cite{Cea1,Cea2,Cea3} with the vector potential alone, the expansions of the action for $n>2$ are excessively overlooked, and then, only the coupling $\chi_{13}H_p\delta|\Delta|$ between the pump effect and Higgs mode in $n=2$ perturbation expansion is left, leading to a zero Higgs-mode generation at clean limit due to $\chi_{13}=0$. But in fact, in $n=3$ expansion, there exists the essential coupling $\chi_{001}H_d^2\delta|\Delta|$ between the second order of the vector-potential drive effect and the Higgs mode, which leads to the finite Higgs-mode generation in Eq.~(\ref{HMEQP}). Previous calculations in Refs.~\onlinecite{Cea1,Cea2,Cea3} that overlooked $n=3$ expansion therefore missed the finite Higgs-mode generation in the second-order response at clean limit.
 
It is also noted that the previous works\cite{Cea1,Cea2,Cea3} with the vector potential alone overlooked the drive effect of the scalar potential. Whereas the finite contribution from this effect to the Higgs-mode generation, which has obtained from GIKE in Eq.~(\ref{shiggs1}), is exactly recovered here in Eq.~(\ref{HMEQP}), as we handle the path-integral approach within the gauge-invariant manner in the present work. As mentioned in Sec.~\ref{sec-GIKE}, this contribution is finite upon cooling to $T=0$, in contrast to the one of the vector potential which that emerges only at finite temperature\cite{DS1}. Whereas because of the gauge structure in superconductors, the consideration of the contribution from the scalar potential is essential here as we have chosen zero {\small $\partial_{\mu}\delta\theta^{(1)}$}.

From the action in Eq.~(\ref{Fac}), after the integration out the Higgs mode, one obtains the finite contribution from the Higgs mode to the third-harmonic current:
\begin{equation}
S[A^4_{\mu}]|_{\rm Higgs}=4\Delta^2_0\frac{e^4v_F^4}{9}\frac{\Big[\lambda_p{A^2_0}\!+\!\gamma_p\Big(\frac{E_{\phi}}{i\Omega}\Big)^2\Big]^2}{(4\Delta_0^2\!-\!4\Omega^2)\beta_p},
\end{equation}
which shows an isotropic signal and exhibits a resonance when $2\Omega=2\Delta_0$, in consistency with the experimental findings\cite{NL8,NL9,NL10}.

\subsubsection{Vanishing charge-density fluctuation}

By using the action in Eq.~(\ref{Fac}), we next discuss the phase mode and charge-density fluctuation in the second-order optical response and its contribution to the third-harmonic current. In the action, for the part that is related to the phase mode and charge-density fluctuation [i.e., the second term in Eq.~(\ref{Fac})], after the integration out the phase mode, one has
\begin{eqnarray}
  &&\!\!\!\!\!\!S[A^4_{\mu}]|_{\rm phase}=-\sum_Q\Big[\Big|\frac{e^2A_0^2}{2m}\!+\!\mu_H\!-\!\frac{\zeta_1eA_0eE_{\phi}}{3mi\Omega}\!+\!\frac{\zeta_2e^2E_{\phi}^2}{3m\Omega^2}\Big|^2\nonumber\\
    &&\mbox{}\!\!\!\!\!\times\Big(\chi_{33}-\frac{\chi_{33}^2}{\chi_{33}}\Big)\!+\!\frac{|\mu_H|^2}{2V_Q}\Big]=-\sum_Q\frac{|\mu_H|^2}{2V_Q}. 
\end{eqnarray}
In above action, there is no coupling term between the charge-density fluctuation and second-order optical fields. Consequently, there is no generation of the charge-density fluctuation in the second-order response to contribute to the third-harmonic current, exactly same as the result [Eq.~(\ref{scdf1})] from the GIKE\cite{GIKE2} and in consistency with the analysis based on the inversion symmetry and charge conservation as mentioned in the introduction. 

Particularly, with $\mu_{H}=0$, from the action in Eq.~(\ref{Fac}), the equation of motion of the phase mode is written as
\begin{equation}\label{spisphase}
-i\Omega\delta\theta^{(2)}=\frac{e^2A_0^2}{2m}\!-\!\frac{\zeta_1eA_0eE_{\phi}}{3mi\Omega}\!+\!\frac{\zeta_2e^2E_{\phi}^2}{3m\Omega^2}.  
\end{equation}
On the right-hand side of above equation, the pump effect $H_p$ of the vector potential (the first term) from $n=2$ perturbation expansion and the second order of the drive effect $H_{\rm LM}^{(1)}$ (the last two terms) from $n=3$ perturbation expansion provide the source terms. Then, one finds a finite second-order response of the phase mode, contributed by both pump and drive effects, in agreement with the result [Eq.~(\ref{sphase1})] from the GIKE\cite{GIKE2}. 

It is noted that in the previous works by Cea {\em et al.}\cite{Cea1,Cea2,Cea3}, the excitation of the phase mode in the second-order response is overlooked. Then, from the second term in Eq.~(\ref{Fac}), one obtains a finite coupling between the second-order optical fields and the charge-density fluctuation $\mu_H$, leading to the finite charge-density fluctuation in the second-order response. Whereas this result violates the analysis based on the inversion symmetry and charge conservation mentioned in the introduction. In fact, by calculating the amplitude and phase modes on an equal footing through the path-integral approach,  the present work shows that the phase mode $\delta\theta^{(2)}$ is excited in the second-order response. Substituting the phase-mode generation in Eq.~(\ref{spisphase}) into the action in Eq.~(\ref{Fac}), one can immediately finds that the charge-density-fluctuation part becomes
\begin{equation}
S[A^4_{\mu}]|_{\rm CDF}=-\sum_Q\Big(\chi_{33}|\mu_H|^2\!+\!\frac{|\mu_H|^2}{2V_Q}\Big).  
\end{equation}
Consequently, the phase-mode generation $\delta\theta^{(2)}$ provides an effective field to exactly cancel the unphysical excitation of the charge-density fluctuation reported by Cea {\em et al.}\cite{Cea1,Cea2,Cea3}, guaranteeing the charge conservation. 

\section{Discussion on Matsubara formalism in derivation of superconducting gap dynamics}
\label{sec-discuss}

In this part, we address a specail issue in the application of Matsubara formalism in superconductors. We show that in the derivation of the superconducting gap dynamics, treating the optical frequency $i\Omega$ as bosonic Matsubara frequencies leads to results against Ginzburg-Landau equation. One has to take $\Omega$ as continuous variable in order to recover/derive Ginzburg-Landau equation.  

Specifically, with the vector potential alone,  either from the phenomenological time-dependent Ginzburg-Landau superconducting Lagrangian [Eq.~(\ref{GLLE})] or through the microscopic Eilenberger equation (Sec.~\ref{sec-Eie}) and path-integral (Sec.~\ref{sec-PIA}) as well as gauge-invariant kinetic equation (Sec.~\ref{sec-GIKE}) approaches, the Higgs-mode dynamics in the second-order response reads:
\begin{equation}
\beta_{H}(4\Delta_0^2-4\Omega^2)\delta|\Delta|=-2\Delta_0\frac{{\lambda}k_F^2}{3m}\frac{e^2A^2}{m},   
\end{equation}
where the response coefficients $\lambda$ from the Ginzburg-Landau theory, Eilenberger equation and path-integral as well as gauge-invariant kinetic equation approaches are given by $\lambda_l=2m\lambda_L/k_F^2={7R(3)}/{(2\pi{T})^2}$ as well as $\lambda_E$ [Eq.~(\ref{DE})], $\lambda_p$ [Eq.~(\ref{lp})] and $\lambda_g$ [Eq.~(\ref{lg1})], respectively.  The Landau parameter $\lambda_L$ was derived by Gorkov near $T_c$ from basic Gorkov equation\cite{G1}. Whereas as mentioned in Secs.~\ref{sec-Eie} and~\ref{sec-PIA}, $\lambda_E$ in Eq.~(\ref{DE}) has been exactly derived by Silaev in Ref.~\onlinecite{Silaev}, but $\lambda_p$ was directly missed in the previous works by Cea {\em et al.}\cite{Cea1,Cea2,Cea3}.

Particularly, using the fact in Eq.~(\ref{fact}), one finds that $\lambda_E$ [Eq.~(\ref{DE})] from Eilenberger equation is directly equivalent to $\lambda_p$ [Eq.~(\ref{lp})] from path-integral approach in Matsubara formalism. However, in Ref.~\onlinecite{Silaev}, $\lambda_E$ derived from Eilenberger equation is considered as zero by taking the optical frequency $i\Omega$ as bosonic Matsubara frequencies $i\Omega_m$, and for bosonic Matsubara frequencies $i\Omega_m$,  $\lambda_p$ in Eq.~(\ref{lp}) derived from path-integral approach also vanishes. This treatment about the optical frequency is indeed the conventional one applied in conductivity calculation of normal metals. Nevertheless, in the derivation of the superconducting gap dynamics here, the vanishing $\lambda_E$ and $\lambda_p$ are strongly against the finite $\lambda_l$ from Ginzburg-Landau theory. Actually, it can be easily demonstrated that both the Eilenberger equation\cite{Ba20} and path-integral approach at the stationary case ($\Omega=0$) can exactly recover the Ginzburg-Landau equation and derive the Ginzburg-Landau kinetic term $\frac{\lambda_L({\nabla}-2ie{\bf A})^2\Delta}{4m}$ (the detailed derivation is given in Appendixes~\ref{EI2GL} and~\ref{PI2GL} for the sake of completeness). In other words, due to the treatment of taking 
$i\Omega$ as bosonic Matsubara frequencies, which eliminates the response coefficient $\lambda$ and hence Ginzburg-Landau kinetic term at finite $\Omega$, an unphysical abrupt change between $\Omega=0$ and $\Omega\rightarrow0$ emerges. This demonstrates that the application of Matsubara formalism in superconductors should be carefully examined, since the treatment that fails to recover the Ginzburg-Landau equation in conventional superconductors can not be correct.

In fact, only by taking $\Omega$ as continuous variable in this circumstance, can one recover/derive Ginzburg-Landau equation. Similarly, only with the continuous $\Omega$, the coefficients $\lambda_E$ [Eq.~(\ref{DE})] from Eilenberger equation and $\lambda_p$ [Eq.~(\ref{lp})] from path-integral approach can exactly recover the finite $\lambda_g$ from gauge-invariant kinetic equation and $\lambda_l$ from Ginzburg-Landau theory at low frequency, as demonstrated in Secs.~\ref{sec-Eie} and~\ref{sec-PIA}. It is noted that the gauge-invariant kinetic equation is developed by using Keldysh Green function approach, irrelevant to the Matsubara formalism, and the coefficient $\lambda_g$ [Eq.~(\ref{lg1})] from this approach can naturally recover the one from Ginzburg-Landau equation\cite{GIKE1}.

The finite second-order response of Higgs mode can also be understood as follows. Specifically, with the vector potential alone at low frequency, it is established that the vector potential drives the Doppler shift\cite{FF4,FF5,FF6} to influence the gap equation:
\begin{equation}
  {\Delta}=g\!\sum_{\bf k}\Delta\frac{f(-{\bf v}_{\bf k}\cdot{e}{\bf A}-E_{\bf k})-f(-{\bf v}_{\bf k}\cdot{e}{\bf A}+E_{\bf k})}{2E_{\bf k}},
\end{equation}
which can be directly derived according to the Hamiltonian in Eq.~(\ref{BdG}). Then, considering the gap variation and weak field, the above equation directly becomes
\begin{equation}
4\Delta_0^2\delta|\Delta|\sum_{\bf k}\frac{\partial_{E_{\bf k}}[\frac{2f(E_{\bf k})-1}{2E_{\bf k}}]}{2E_{\bf k}}\!=\!-2\Delta_0\frac{v_F^2e^2A^2\sum_{\bf k}\frac{\partial^2_{E_{\bf k}}f(E_{\bf k})}{2E_{\bf k}}}{3},
\end{equation}
which is exactly same as the ones derived from gauge-invariant kinetic equation [Eq.~(\ref{shiggs1})] and path-integral approach [Eq.~(\ref{HMEQP})] at low continuous optical frequency.

\section{Summary}

In summary, through three different microscopic approaches including the GIKE, Eilenberger equation as well as path-integral approach, the present study arrives at unified conclusion about the finite Higgs-mode generation in the second-order optical response of superconductors at clean limit, in consistency with the phenomenological Ginzburg-Landau theory. Moreover, while the density-related effect is hard to handle in the quasiclassical Eilenberger equation, the vanishing charge-density fluctuation in the second-order response, which agrees with the charge conservation, is obtained within the GIKE and path-integral approach. Consequently, the present work resolves the controversies among various studies in the literature (whether the experimentally observed third-harmonic current\cite{NL7,NL8,NL9,NL10,NL11,DHM2,DHM3} is attributed to the Higgs-mode\cite{NL10,NL11,DHM2,DHM3,FHM,GIKE2} or charge-density-fluctuation\cite{Cea1,Cea2,Cea3} generation, and whether one has to rely on impurity to explain the experimentally observed Higgs-mode generation\cite{Cea1,Cea2,Cea3,GIKE2,Silaev,FHM,Am16,ImR1,ImR2,ImR3}), and can therefore help understanding the experimental findings. 

Specifically, with the vector potential alone, by separately using the GIKE as well as Eilenberger equation and path-integral approach, we obtain the exactly same finite Higgs-mode generation in the second-order optical response at clean limit.  This finite Higgs-mode generation is solely contributed by the drive effect ${\bf p}\cdot{e{\bf A}}/m$ of the vector potential, and near $T_c$, exactly recovers the one from the phenomenological Ginzburg-Landau theory. Nevertheless, the previous works within the path-integral approach by Cea {\em et al.}\cite{Cea1,Cea2,Cea3} and Eilenberger equation by Silaev\cite{Silaev} missed this finite generation because of the flaws in their derivations. A disscussion about the application of Matsubara formalism in the derivation of superconducting gap dynamics is given, and it is demonstrated that taking the optical frequency as continuous variable is essential to recover/derive the Ginzburg-Landau equation.

According to the gauge structure of superconductors\cite{gi0}, among the scalar potential $\phi$, vector potential ${\bf A}$ as well as phase-related effective electromagnetic potential  {\small $\partial_{\mu}\delta\theta$}, one can not choose two quantities simultaneously to be zero, e.g., considering the vector potential alone. We therefore extend the path-integral approach to include electromagnetic effects from the scalar potential and phase mode. Then, in the second-order response at clean limit,  a finite contribution in the Higgs-mode generation from the drive effect of scalar potential as well as the vanishing charge-density fluctuation, which have previously been obtained from GIKE\cite{GIKE2}, are exactly recovered.

The contribution of the scalar potential is finite upon cooling to $T=0$,  differing from the one of the vector potential that emerges only at finite temperature\cite{DS1}. This difference is due to the fact that the superconductors can directly respond to vector potential ${\bf A}$ (Meissner effect/Ginzburg-Landau kinetic term) in addition to the electric field ${\bf E}=-{\nabla}_{\bf R}\phi-\partial_t{\bf A}$ (optical-electric-field effect), differing from normal metals that solely respond to ${\bf E}$.
Consequently, in contrast to the scalar potential that only captures the optical-electric-field effect,  the vector potential also characterizes the Meissner effect/Ginzburg-Landau kinetic term in addition to the optical-electric-field effect. Particularly, because of the gauge structure of superconductors\cite{gi0}, the inclusion of the contribution from the scalar potential here is essential, since we have chosen zero {\small $\partial_{\mu}\delta\theta^{(l)}$} (i.e., a spatially uniform background phase mode $\delta\theta^{(l)}$).

Although the uniform background phase mode $\delta\theta^{(l)}$ does not manifest itself in the measurable optical properties, the inclusion of this mode is essential for theoretical description to cancel the unphysical effect, as pointed out by Nambu in his Nobel lecture\cite{gi1}. The linear response of the background phase mode $\delta\theta^{(1)}$ cancels the unphysical longitudinal vector potential in ${\bf p}_s$, as established in the literature\cite{AK,AK2,Ba0,pm0,Am0,Ba9,Ba10,pm5,pi1,pi4,GIKE2}. Then, the superconducting momentum ${\bf p}_s$ that appears in the previous theoretical descriptions such as Ginzburg-Landau equation\cite{G1} and Meissner supercurrent\cite{G1} as well as Anderson-pump effect\cite{Am1,Am2,Am3,Am4,Am7,Am8,Am9,Am10,Am11,Am14,NL7,NL8,NL9,NL10,NL11} only involves the physical transverse vector potential. The previous works by Cea {\em et al.}\cite{Cea1,Cea2,Cea3} overlooked the phase mode, and obtained the excited charge-density fluctuation in the second-order optical response of superconductors. Whereas this result in systems with the inversion symmetry violates the law of charge conservation. We show in the present work that a background phase mode
 $\delta\theta^{(2)}$ is actually generated in the second-order response, and exactly cancel the unphysical excitation of the charge-density fluctuation reported in Refs.~\onlinecite{Cea1,Cea2,Cea3}, guaranteeing the charge conservation.\\

\begin{acknowledgments}
The authors acknowledge financial support from
the National Natural Science Foundation of 
China under Grants No.\ 11334014 and No.\ 61411136001.  
\end{acknowledgments}

\begin{widetext}

\begin{appendix}

\section{Derivation of correlation coefficients}
\label{ADCC}

In this part, from Eqs.~(\ref{chiij})-(\ref{chii3k}), we present the specific expressions of the related correlation coefficients at low frequency ($Q_0<E_{\bf k}$) and long-wave limit (${\bf Q}\rightarrow0$):
\begin{eqnarray}
  \chi_{11}\!-\!U^{-1}\!\!&=&\!\!\frac{1}{2}\sum_{p}{\rm Tr}\Big[\frac{p_0\!+\!2Q_0\!+\!\xi_{\bf k}\tau_3\!+\!\Delta_0\tau_1}{(p_0+2Q_0)^2-E^2_{\bf k}}\tau_1\frac{p_0\!+\!\xi_{\bf k}\tau_3\!+\!\Delta_0\tau_1}{p_0^2-E^2_{\bf k}}\tau_1\Big]\!-\!U^{-1}=\sum_p\frac{(p_0+2Q_0)p_0+\Delta_0^2-\xi^2_{\bf k}}{[(p_0+2Q_0)^2-E^2_{\bf k}](p_0^2-E^2_{\bf k})}-U^{-1}\nonumber\\
  \!\!&=&\!\!\sum_p\Big\{\frac{4\Delta_0^2-(p_0+2Q_0-p_0)^2}{2[(p_0+2Q_0)^2-E^2_{\bf k}](p_0^2-E^2_{\bf k})}+\frac{1}{2(p_0^2-E^2_{\bf k})}+\frac{1}{2[(p_0+2Q_0)^2-E^2_{\bf k}]}\Big\}-\sum_p\frac{1}{p_0^2-E^2_{\bf k}}\nonumber\\
  \!\!&\approx&\!\!\sum_p\frac{4\Delta_0^2\!-\!4Q_0^2}{2(p_0^2-E^2_{\bf k})^2}\!=\!4({\Delta_0^2\!-\!Q_0^2})\sum_{\bf k}\frac{1}{2E_{\bf k}^2}\Big[\frac{1\!-\!2f(E_{\bf k})}{2E_{\bf k}}\!+\!\partial_{E_{\bf k}}f(E_{\bf k})\Big]=4({\Delta_0^2\!-\!Q_0^2})\sum_{\bf k}\frac{\partial_{E_{\bf k}}}{2E_{\bf k}}\Big[\frac{2f(E_{\bf k})\!-\!1}{2E_{\bf k}}\Big],~~~~~ \label{c11}
\end{eqnarray}
\begin{eqnarray}
  \chi_{33}&=&\frac{1}{2}\sum_{p}{\rm Tr}\Big[\frac{p_0\!+\!2Q_0\!+\!\xi_{\bf k}\tau_3\!+\!\Delta_0\tau_1}{(p_0+2Q_0)^2-E^2_{\bf k}}\tau_3\frac{p_0\!+\!\xi_{\bf k}\tau_3\!+\!\Delta_0\tau_1}{p_0^2-E^2_{\bf k}}\tau_3\Big]=\sum_p\frac{(p_0\!+\!2Q_0)p_0\!+\!\xi^2_{\bf k}\!-\!\Delta_0^2}{[(p_0+2Q_0)^2-E^2_{\bf k}](p_0^2-E^2_{\bf k})}\nonumber\\
  &=&\sum_{\bf k}\Big\{\Big[\frac{f(E_{\bf k})}{2E_{\bf k}}\frac{2\xi_{\bf k}^2\!+\!2Q_0E_{\bf k}}{4Q_0(E_{\bf k}\!+\!Q_0)}\!-\!\frac{f(E_{\bf k}\!-\!2Q_0)}{2E_{\bf k}}\frac{2\xi_{\bf k}^2\!-\!2Q_0E_{\bf k}}{4Q_0(E_{\bf k}\!-\!Q_0)}\Big]\!+\![E_{\bf k}\!\rightarrow\!-E_{\bf k}]\Big\}\approx\sum_{\bf k}\frac{\Delta^2_0}{E_{\bf k}^2\!-\!Q_0^2}\frac{2f(E_{\bf k})\!-\!1}{2E_{\bf k}},  \label{c33}
\end{eqnarray}
\begin{equation}
  \chi_{13}=\frac{1}{2}\sum_{p}{\rm Tr}\Big[\frac{p_0\!+\!2Q_0\!+\!\xi_{\bf k}\tau_3\!+\!\Delta_0\tau_1}{(p_0+2Q_0)^2-E^2_{\bf k}}\tau_1\frac{p_0\!+\!\xi_{\bf k}\tau_3\!+\!\Delta_0\tau_1}{p_0^2-E^2_{\bf k}}\tau_3\Big]=\sum_p\frac{2\xi_{\bf k}\Delta_0}{[(p_0+2Q_0)^2-E^2_{\bf k}](p_0^2-E^2_{\bf k})}=0,\label{c13}
\end{equation}
\begin{eqnarray}
  \chi_{003}&=&\sum_{p}{\rm Tr}\Big[\frac{p_0\!+\!2Q_0\!+\!\xi_{\bf k}\tau_3\!+\!\Delta_0\tau_1}{(p_0+2Q_0)^2-E^2_{\bf k}}\frac{p_0\!+\!Q_0\!+\!\xi_{\bf k}\tau_3\!+\!\Delta_0\tau_1}{(p_0+Q_0)^2-E^2_{\bf k}}\frac{p_0\!+\!\xi_{\bf k}\tau_3\!+\!\Delta_0\tau_1}{p_0^2-E^2_{\bf k}}\tau_3\Big]\nonumber\\
  &=&\sum_{p}2\xi_{\bf k}\frac{(p_0+2Q_0)(p_0+Q_0)+\xi_{\bf k}^2+\Delta_0^2+2p^2_0+3p_0Q_0)}{[(p_0+2Q_0)^2-E^2_{\bf k}][(p_0+Q_0)^2-E^2_{\bf k}](p_0^2-E^2_{\bf k})}=0. \label{c003}
\end{eqnarray}  
\begin{eqnarray}
  \chi_{0{\bar 3}3}&=&\sum_{p}{\rm Tr}\Big\{\frac{p_0\!+\!2Q_0\!+\!\xi_{\bf k}\tau_3\!+\!\Delta_0\tau_1}{(p_0+2Q_0)^2-E^2_{\bf k}}\partial_{\xi_{\bf k}}\Big[\frac{p_0\!+\!Q_0\!+\!\xi_{\bf k}\tau_3\!+\!\Delta_0\tau_1}{(p_0+Q_0)^2-E^2_{\bf k}}\Big]\tau_3\frac{p_0\!+\!\xi_{\bf k}\tau_3\!+\!\Delta_0\tau_1}{p_0^2-E^2_{\bf k}}\tau_3\Big\}\nonumber\\
  &=&\sum_{p}2\Big\{\frac{2\xi_{\bf k}(p_0+Q_0)}{[(p_0\!+\!2Q_0)^2\!-\!E^2_{\bf k}](p_0^2\!-\!E^2_{\bf k})}\!+\!\frac{[(p_0\!+\!2Q_0)(p_0\!+\!Q_0)\!+\!E_{\bf k}^2]p_0\!+\!(2p_0\!+\!3Q_0)(\xi_{\bf k}^2-\Delta_0^2)}{[(p_0\!+\!2Q_0)^2\!-\!E^2_{\bf k}](p_0^2\!-\!E^2_{\bf k})}\partial_{\xi_{\bf k}}\Big\}\Big[\frac{1}{(p_0\!+\!Q_0)^2\!-\!E^2_{\bf k}}\Big]\nonumber\\
  &=&0,~~~~\label{c0b33}
\end{eqnarray}
\begin{eqnarray}
  \chi_{3{\bar 0}3}&=&\sum_{p}{\rm Tr}\Big\{\frac{p_0\!+\!2Q_0\!+\!\xi_{\bf k}\tau_3\!+\!\Delta_0\tau_1}{(p_0+2Q_0)^2-E^2_{\bf k}}\tau_3\partial_{\xi_{\bf k}}\Big[\frac{p_0\!+\!Q_0\!+\!\xi_{\bf k}\tau_3\!+\!\Delta_0\tau_1}{(p_0+Q_0)^2-E^2_{\bf k}}\Big]\frac{p_0\!+\!\xi_{\bf k}\tau_3\!+\!\Delta_0\tau_1}{p_0^2-E^2_{\bf k}}\tau_3\Big\}\nonumber\\
  &=&\sum_{p}2\Big\{\frac{2\xi_{\bf k}(p_0+Q_0)}{[(p_0\!+\!2Q_0)^2\!-\!E^2_{\bf k}](p_0^2\!-\!E^2_{\bf k})}\!+\!\frac{(p_0\!+\!2Q_0)[(p_0\!+\!Q_0)p_0\!+\!E_{\bf k}^2]\!+\!(2p_0\!+\!Q_0)(\xi_{\bf k}^2-\Delta_0^2)}{[(p_0\!+\!2Q_0)^2\!-\!E^2_{\bf k}](p_0^2\!-\!E^2_{\bf k})}\partial_{\xi_{\bf k}}\Big\}\Big[\frac{1}{(p_0\!+\!Q_0)^2\!-\!E^2_{\bf k}}\Big]\nonumber\\ 
  &=&0,~~~~\label{c3b03}
\end{eqnarray}
\begin{eqnarray}
  \chi_{3{\bar{\bar 3}}3}&=&\sum_{p}{\rm Tr}\Big\{\frac{p_0\!+\!2Q_0\!+\!\xi_{\bf k}\tau_3\!+\!\Delta_0\tau_1}{(p_0+2Q_0)^2-E^2_{\bf k}}\tau_3\partial^2_{\xi_{\bf k}}\Big[\frac{p_0\!+\!Q_0\!+\!\xi_{\bf k}\tau_3\!+\!\Delta_0\tau_1}{(p_0+Q_0)^2-E^2_{\bf k}}\Big]\tau_3\frac{p_0\!+\!\xi_{\bf k}\tau_3\!+\!\Delta_0\tau_1}{p_0^2-E^2_{\bf k}}\tau_3\Big\}\nonumber\\
  &=&\sum_{p}2\Big\{\frac{2[(p_0\!+\!2Q_0)p_0\!+\!\xi^2_{\bf k}\!-\!\Delta_0^2]\partial_{\xi_{\bf k}}}{[(p_0\!+\!2Q_0)^2\!-\!E^2_{\bf k}](p_0^2\!-\!E^2_{\bf k})}\!+\!\frac{\xi_{\bf k}[(p_0\!+\!2Q_0)(p_0\!+\!Q_0)\!+\!\xi_{\bf k}^2\!-\!3\Delta_0^2\!+\!(2p_0\!+\!3Q_0)p_0]\partial^2_{\xi_{\bf k}}}{[(p_0+2Q_0)^2-E^2_{\bf k}](p_0^2-E^2_{\bf k})}\Big\}\Big[\frac{1}{(p_0\!+\!Q_0)^2\!-\!E^2_{\bf k}}\Big]\nonumber\\
  &=&0, \label{c3b33}
\end{eqnarray}
\begin{eqnarray}
  \chi_{001}&=&\sum_{p}{\rm Tr}\Big\{\frac{p_0\!+\!2Q_0\!+\!\xi_{\bf k}\tau_3\!+\!\Delta_0\tau_1}{(p_0+2Q_0)^2-E^2_{\bf k}}\frac{p_0\!+\!Q_0\!+\!\xi_{\bf k}\tau_3\!+\!\Delta_0\tau_1}{(p_0+Q_0)^2-E^2_{\bf k}}\frac{p_0\!+\!\xi_{\bf k}\tau_3\!+\!\Delta_0\tau_1}{p_0^2-E^2_{\bf k}}\tau_1\Big\}\nonumber\\
  &=&\sum_{p}2\Delta_0\frac{E^2_{\bf k}\!+\!(p_0\!+\!2Q_0)p_0\!+\!(p_0\!+\!Q_0)p_0\!+\!(p_0\!+\!2Q_0)(p_0\!+\!Q_0)}{[(p_0+2Q_0)^2-E^2_{\bf k}][(p_0+Q_0)^2-E^2_{\bf k}](p_0^2-E^2_{\bf k})}\nonumber\\
  &=&\sum_{\bf k}\Big\{\Big[\frac{f(E_{\bf k}-2Q_0)\delta_{ip_0+2Q_0,E_{\bf k}}}{[(ip_0\!+\!Q_0)^2\!-\!E^2_{\bf k}][(ip_0)^2\!-\!E^2_{\bf k}]}\!+\!\frac{f(E_{\bf k}-Q_0)\delta_{ip_0+Q_0,E_{\bf k}}}{[(ip_0\!+\!2Q_0)^2\!-\!E^2_{\bf k}][(ip_0)^2\!-\!E^2_{\bf k}]}\!+\!\frac{f(E_{\bf k})\delta_{ip_0,E_{\bf k}}}{[(ip_0\!+\!2Q_0)^2\!-\!E^2_{\bf k}][(ip_0\!+\!Q_0)^2\!-\!E^2_{\bf k}]}\Big]\nonumber\\
  &&\mbox{}\times\frac{2\Delta_0[E^2_{\bf k}\!+\!(ip_0\!+\!2Q_0)ip_0\!+\!(ip_0\!+\!Q_0)ip_0\!+\!(ip_0\!+\!2Q_0)(ip_0\!+\!Q_0)]}{2E_{\bf k}}+\{E_{\bf k}\rightarrow-E_{\bf k}\}\Big\}  \nonumber\\
  &=&2\Delta_0\sum_{\bf k}\Big\{\Big[\frac{f(E_{\bf k}\!-\!2Q_0)(ip_0\!+\!Q_0\!+\!ip_0\!+\!2Q_0)(ip_0\!+\!ip_0\!+\!2Q_0)\delta_{ip_0+2Q_0,E_{\bf k}}}{2E_{\bf k}[(ip_0\!+\!Q_0)^2\!-\!(ip_0\!+\!2Q_0)^2][(ip_0)^2\!-\!(ip_0\!+\!2Q_0)^2]}\!+\!\frac{f(E_{\bf k}\!-\!Q_0)(ip_0\!+\!2Q_0\!+\!ip_0\!+\!Q_0)}{2E_{\bf k}[(ip_0\!+\!2Q_0)^2\!-\!(ip_0\!+\!Q_0)^2]}\nonumber\\
    &&\mbox{}\times\frac{(ip_0\!+\!Q_0\!+\!ip_0)\delta_{ip_0+Q_0,E_{\bf k}}}{[(ip_0)^2\!-\!(ip_0\!+\!Q_0)^2]}\!+\!\frac{f(E_{\bf k})\delta_{ip_0,E_{\bf k}}(ip_0\!+\!2Q_0\!+\!ip_0)(ip_0\!+\!Q_0\!+\!ip_0)}{2E_{\bf k}[(ip_0\!+\!2Q_0)^2\!-\!(ip_0)^2][(ip_0\!+\!Q_0)^2\!-\!(ip_0)^2]}\Big]+[E_{\bf k}\rightarrow-E_{\bf k}]\Big\} \nonumber\\
  &=&\frac{2\Delta_0}{2Q_0^2}\sum_{\bf k}\Big\{\Big[\frac{f(E_{\bf k}\!-\!2Q_0)}{2E_{\bf k}}\!+\!\frac{f(E_{\bf k})}{2E_{\bf k}}\!-\!\frac{2f(E_{\bf k}\!-\!Q_0)}{2E_{\bf k}}\Big]+\{E_{\bf k}\rightarrow-E_{\bf k}\}\Big\}. \label{c001}
\end{eqnarray}
\begin{eqnarray}
  \chi_{3{\bar{\bar 3}}1}&=&\frac{1}{3}\sum_{p}{\rm Tr}\Big\{\frac{p_0\!+\!2Q_0\!+\!\xi_{\bf k}\tau_3\!+\!\Delta_0\tau_1}{(p_0+2Q_0)^2-E^2_{\bf k}}\tau_3\partial^2_{\xi_{\bf k}}\Big[\frac{p_0\!+\!Q_0\!+\!\xi_{\bf k}\tau_3\!+\!\Delta_0\tau_1}{(p_0+Q_0)^2-E^2_{\bf k}}\Big]\tau_3\frac{p_0\!+\!\xi_{\bf k}\tau_3\!+\!\Delta_0\tau_1}{p_0^2-E^2_{\bf k}}\tau_1\Big\}\nonumber\\ 
  &=&\frac{1}{3}\sum_{p}\partial_{\xi_{\bf k'}}\Big\{{\rm Tr}\Big[\frac{p_0\!+\!Q_0\!+\!\xi_{\bf k}\tau_3\!+\!\Delta_0\tau_1}{(p_0+Q_0)^2-E^2_{\bf k}}\frac{p_0\!+\!\xi_{\bf k'}\tau_3\!-\!\Delta_0\tau_1}{p_0^2-E^2_{\bf k'}}\frac{p_0\!-\!Q_0\!+\!\xi_{\bf k}\tau_3\!+\!\Delta_0\tau_1}{(p_0-Q_0)^2-E^2_{\bf k}}\tau_1\Big]\Big\}_{\bf k'{\rightarrow}k}\nonumber\\
  &=&\frac{2\Delta_0}{3}\sum_{p}\partial_{\xi_{\bf k'}}\Big\{\frac{p_0^2+Q_0^2+\xi_{\bf k}^2+2\xi_{\bf k}\xi_{\bf k'}-\Delta_0^2}{[(p_0+Q_0)^2-E^2_{\bf k}](p_0^2-E_{\bf k'}^2)[(p_0-Q_0)^2-E^2_{\bf k}]}\Big\}_{\bf k'{\rightarrow}k}\nonumber\\
  &=&\frac{2\Delta_0}{3}\sum_{p}\partial_{\xi_{\bf k'}}\Big\{\frac{(\xi_{\bf k}+\xi_{\bf k'})^2+Q_0^2}{[(p_0+Q_0)^2-E^2_{\bf k}](p_0^2-E_{\bf k'}^2)[(p_0-Q_0)^2-E^2_{\bf k}]}+\frac{1}{[(p_0+Q_0)^2-E^2_{\bf k}][(p_0-Q_0)^2-E^2_{\bf k}]}\Big\}_{\bf k'{\rightarrow}k}\nonumber\\
  &=&\frac{2\Delta_0}{3(iQ_0)^2}\!\sum_{\bf k}\big\{\big[2\!+\!4\xi_{\bf k}\partial_{\xi_{\bf k'}}\!+\!(4\xi_{\bf k}^2\!+\!Q_0^2)\partial_{\xi_{\bf k'}}^2\big]I_{\bf k}(\xi_{\bf k'})\big\}_{\bf k'{\rightarrow}k}\!\approx\!\frac{2\Delta_0}{(iQ_0)^2}\!\sum_{\bf k}\frac{4\xi^2_{\bf k}}{3}\Big\{\frac{3\xi_{\bf k}^2[1\!-\!2f(E_{\bf k})]}{8E^7_{\bf k}}\!+\!\frac{3\xi_{\bf k}^2}{4E_{\bf k}^6}\partial_{E_{\bf k}}f(E_{\bf k})\Big\}\nonumber\\
&=&\frac{2\Delta_0}{(iQ_0)^2}\!\sum_{\bf k}\frac{\xi_{\bf k}^4}{E_{\bf k}^5}\partial_{E_{\bf k}}\Big[\frac{2f(E_{\bf k})-1}{2E_{\bf k}}\Big],
\end{eqnarray}
\begin{eqnarray}
  \chi_{3{\bar 0}1}\!-\!\chi_{0{\bar 3}1}\!\!&=&\!\!\sum_{p}\partial_{\xi_{\bf k'}}\Big\{{\rm Tr}\Big[\frac{p_0\!+\!Q_0\!+\!\xi_{\bf k}\tau_3\!+\!\Delta_0\tau_1}{(p_0+Q_0)^2-E^2_{\bf k}}\frac{[\tau_3,p_0\!+\!\xi_{\bf k'}\tau_3\!+\!\Delta_0\tau_1]}{p_0^2-E^2_{\bf k'}}\frac{p_0\!-\!Q_0\!+\!\xi_{\bf k}\tau_3\!+\!\Delta_0\tau_1}{(p_0-Q_0)^2-E^2_{\bf k}}\tau_1\Big]\Big\}_{\bf k'{\rightarrow}k}\nonumber\\
 ~\!\!&=&\!\!\sum_{p}\partial_{\xi_{\bf k'}}\Big\{\frac{-8Q_0\Delta_0\xi_{\bf k}}{[(p_0\!+\!Q_0)^2\!-\!E^2_{\bf k}](p_0^2\!-\!E_{\bf k'}^2)[(p_0\!-\!Q_0)^2\!-\!E^2_{\bf k}]}\Big\}_{\bf k'{\rightarrow}k}\!\!=\!\frac{8i\Delta_0}{iQ_0}\sum_{\bf k}\xi_{\bf k}\big[\partial_{\xi_{\bf k'}}I_{\bf k}(\xi_{\bf k'})\big]_{\bf k'{\rightarrow}k}\!\!\approx\!0,~~~~~ 
\end{eqnarray}
with
\begin{eqnarray}
  I_{\bf k}(\xi_{\bf k'})&=&\int\frac{dp_0}{2\pi}\frac{(iQ_0)^2}{[(p_0\!+\!Q_0)^2\!-\!E^2_{\bf k}](p_0^2\!-\!E_{\bf k'}^2)[(p_0\!-\!Q_0)^2\!-\!E^2_{\bf k}]}\nonumber\\
  &=&\frac{f(E_{\bf k'})}{2E_{\bf k'}}\frac{1}{(4E_{\bf k'}^2-Q^2_0)}\Big[1\!+\!\frac{\xi_{\bf k}^2\!-\!\xi^2_{\bf k'}}{(2E_{\bf k'}\!+\!Q_0)Q_0}-\frac{\xi_{\bf k}^2\!-\!\xi^2_{\bf k'}}{(2E_{\bf k'}\!-\!Q_0)Q_0}\!+\!\frac{(\xi_{\bf k}^2\!-\!\xi^2_{\bf k'})^2}{(2E_{\bf k'}\!+\!Q_0)^2Q^2_0}\!+\!\frac{(\xi_{\bf k}^2\!-\!\xi^2_{\bf k'})^2}{(2E_{\bf k'}\!-\!Q_0)^2Q^2_0}\nonumber\\
&&\mbox{}\!-\!\frac{(\xi_{\bf k}^2\!-\!\xi^2_{\bf k'})^2}{(4E^2_{\bf k'}\!-\!Q^2_0)Q^2_0}\Big]\!-\!\frac{f(E_{\bf k}\!-\!Q_0)}{4E_{\bf k}}\frac{1}{(2E_{\bf k}\!-\!2Q_0)(2E_{\bf k}\!-\!Q_0)}\Big[1\!+\!\frac{\xi_{\bf k}^2\!-\!\xi^2_{\bf k'}}{(2E_{\bf k}\!-\!Q_0)Q_0}\!+\!\frac{(\xi_{\bf k}^2\!-\!\xi^2_{\bf k'})^2}{(2E_{\bf k}\!-\!Q_0)^2Q^2_0}\Big]\nonumber\\
  &&\mbox{}-\frac{f(E_{\bf k}\!+\!Q_0)}{4E_{\bf k}}\frac{1}{(2E_{\bf k}\!+\!2Q_0)(2E_{\bf k}\!+\!Q_0)}\Big[1\!-\!\frac{\xi_{\bf k}^2\!-\!\xi^2_{\bf k'}}{(2E_{\bf k}\!-\!Q_0)Q_0}\!+\!\frac{(\xi_{\bf k}^2\!-\!\xi^2_{\bf k'})^2}{(2E_{\bf k}\!+\!Q_0)^2Q^2_0}\Big]+\{E_{\bf k}\rightarrow-E_{\bf k},E_{\bf k'}\rightarrow-E_{\bf k'}\}\nonumber\\
&\approx&\frac{f(E_{\bf k'})}{2E_{\bf k'}(4E_{\bf k'}^2-Q^2_0)}\Big[1\!+\!\frac{\xi_{\bf k}^2\!-\!\xi^2_{\bf k'}}{(2E_{\bf k'}\!+\!Q_0)Q_0}-\frac{\xi_{\bf k}^2\!-\!\xi^2_{\bf k'}}{(2E_{\bf k'}\!-\!Q_0)Q_0}\!+\!\frac{(\xi_{\bf k}^2\!-\!\xi^2_{\bf k'})^2}{(2E_{\bf k'}\!+\!Q_0)^2Q^2_0}\!+\!\frac{(\xi_{\bf k}^2\!-\!\xi^2_{\bf k'})^2}{(2E_{\bf k'}\!-\!Q_0)^2Q^2_0}\nonumber\\
&&\mbox{}\!-\!\frac{(\xi_{\bf k}^2\!-\!\xi^2_{\bf k'})^2}{(4E^2_{\bf k'}\!-\!Q^2_0)Q^2_0}\Big]\!-\!\frac{[1\!-\!Q_0\partial_{E_{\bf k}}+\frac{Q_0^2}{2}\partial_{E_{\bf k}}^2]f(E_{\bf k})}{4E_{\bf k}(2E_{\bf k}\!-\!2Q_0)(2E_{\bf k}\!-\!Q_0)}\Big[1\!+\!\frac{\xi_{\bf k}^2\!-\!\xi^2_{\bf k'}}{(2E_{\bf k}\!-\!Q_0)Q_0}\!+\!\frac{(\xi_{\bf k}^2\!-\!\xi^2_{\bf k'})^2}{(2E_{\bf k}\!-\!Q_0)^2Q^2_0}\Big]\nonumber\\
  &&\mbox{}-\frac{[1\!+\!Q_0\partial_{E_{\bf k}}+\frac{Q_0^2}{2}\partial_{E_{\bf k}}^2]f(E_{\bf k})}{4E_{\bf k}(2E_{\bf k}\!+\!2Q_0)(2E_{\bf k}\!+\!Q_0)}\Big[1\!-\!\frac{\xi_{\bf k}^2\!-\!\xi^2_{\bf k'}}{(2E_{\bf k}\!-\!Q_0)Q_0}\!+\!\frac{(\xi_{\bf k}^2\!-\!\xi^2_{\bf k'})^2}{(2E_{\bf k}\!+\!Q_0)^2Q^2_0}\Big]+\{E_{\bf k}\rightarrow-E_{\bf k},E_{\bf k'}\rightarrow-E_{\bf k'}\},  
\end{eqnarray}
and hence
\begin{equation}
[I_{\bf k}(\xi_{\bf k'})]_{{\bf k}\rightarrow{\bf k'},\Omega\ll{E_{\bf k}}}=\Big[\frac{2f(E_{\bf k})-1}{8E^3_{\bf k}}-\frac{2f(E_{\bf k})-1}{16E^3_{\bf k}}-\frac{2f(E_{\bf k})-1}{16E^3_{\bf k}}\Big]=0,  
\end{equation}
\begin{eqnarray}
  [\partial_{\xi_{\bf k'}}I_{\bf k}(\xi_{\bf k'})]_{{\bf k}\rightarrow{\bf k'},\Omega\ll{E_{\bf k}}}&=&\frac{\xi_{\bf k}}{Q_0(2E_{\bf k})^4}\Big[\frac{2Q_0f(E_{\bf k})}{E_{\bf k}}\!+\!f(E_{\bf k})\Big(1\!+\!\frac{2Q_0}{E_{\bf k}}\Big)\!-\!Q_0\partial_{E_{\bf k}}f(E_{\bf k})-f(E_{\bf k})\Big(1\!-\!\frac{2Q_0}{E_{\bf k}}\Big)\!-\!Q_0\partial_{E_{\bf k}}f(E_{\bf k})\Big]\nonumber\\
  &&+\frac{\xi_{\bf k}}{E_{\bf k}}\partial_{E_{\bf k}}\Big[\frac{f(E_{\bf k})}{8E_{\bf k}^3}\Big]+\{E_{\bf k}\rightarrow-E_{\bf k}\}=0,  
\end{eqnarray}
\begin{eqnarray}
  [\partial_{\xi_{\bf k'}}^2I_{\bf k}(\xi_{\bf k'})]_{{\bf k}\rightarrow{\bf k'},\Omega\ll{E_{\bf k}}}&=&\frac{1}{\xi_{\bf k}}\partial_{\xi_{\bf k}}\Big[\frac{f(E_{\bf k})}{8E_{\bf k}^3}\Big]\!-\!\partial_{\xi_{\bf k}}^2\Big[\frac{f(E_{\bf k})}{8E_{\bf k}^3}\Big]\!-\!\xi_{\bf k}\partial_{\xi_{\bf k}}\Big[\frac{f(E_{\bf k})}{4E_{\bf k}^5}\Big]\!+\!\frac{\xi_{\bf k}^2f(E_{\bf k})}{4E_{\bf k}^7}\!-\!\frac{\xi_{\bf k}^2}{4E_{\bf k}^6}\partial_{E_{\bf k}}f(E_{\bf k})\!+\!\frac{\xi_{\bf k}^2}{8E_{\bf k}^5}\partial^2_{E_{\bf k}}f(E_{\bf k})\nonumber\\
  &&\mbox{}+\{E_{\bf k}\rightarrow-E_{\bf k}\}=\frac{3\xi_{\bf k}^2[1-2f(E_{\bf k})]}{8E^7_{\bf k}}+\frac{3\xi_{\bf k}^2}{4E_{\bf k}^6}\partial_{E_{\bf k}}f(E_{\bf k}).
\end{eqnarray}
Here, we have applied the Wick rotation [i.e., $\int\frac{dp_0}{2\pi}F(p_0)\rightarrow{T}\sum_{\omega_n}F(i\omega_n)$] to map the frequency integral into the Matsubara frequency summation\cite{G1}. It is noted that $\chi_{13}$ [Eq.~(\ref{c13})], $\chi_{003}$ [Eq.~(\ref{c003})], $\chi_{0{\bar 3}3}$ [Eq.~(\ref{c0b33})], $\chi_{3{\bar 0}3}$ [Eq.~(\ref{c3b03})] and $\chi_{3{\bar{\bar 3}}3}$ [Eq.~(\ref{c3b33})] vanish as the consequence of the particle-hole symmetry, which eliminates the terms with the odd order of $\xi_{\bf k}$ in the summation of ${\bf k}$. The second-order correlation coefficients $\chi_{11}$ [Eq.~(\ref{c11})] and $\chi_{33}$ [Eq.~(\ref{c33})] and $\chi_{13}$ [Eq.~(\ref{c13})] here are exactly same as the ones obtained in the previous works by Cea {\em et al.}\cite{Cea1,Cea2,Cea3}. Moreover, one also finds that the phase-related coefficient $\chi_{33}$ [Eq.~(\ref{c33})] from the path-integral approach is exactly same as the one $u_g$ [Eq.~(\ref{ug})] from GIKE.\\

\section{Derivation of Ginzburg-Landau equation from Eilenberger equation}
\label{EI2GL}

In this part, we present the derivation of Ginzburg-Landau equation from Eilenberger equation at clean and stationary case\cite{Ba20}. In this circumstance, the Eilenberger equation in Matsubara formalism reads
\begin{equation}
[i\omega_n\tau_3-{\hat \Delta}({\bf R})\tau_3,g]\!+\!i{\bf v}_F\cdot{\bf \nabla}_{\bf R}g\!+\!{e{\bf A}\cdot{\bf v}_F}[\tau_3,g]=0.\label{Ei2GL_1}
\end{equation}  
At the weak field, the quasiclassical $\tau_3$-Green function can be expanded as $g=g^{(0)}+\sum_{n=1}\delta{g}^{(n)}$, with $\delta{g}^{(n)}$ being the $n$-th order response. 

Considering the anomalous Green function (i.e., off-diagonal part $\delta{g}^{(n)}_{12}$) of $\delta{g}^{(n)}$, from Eq.~(\ref{Ei2GL_1}), one has  
\begin{equation}
  2i\omega_n\delta{g}^{(n)}_{12}=-i{\bf v}_F\cdot(\partial_{\bf R}-2ie{\bf A})\delta{g}^{(n-1)}_{12}.
\end{equation}
Then, keeping the expansions up to the second-order response, one finds the solution:
\begin{equation}
g_{12}=g^{(0)}_{12}-\frac{{\bf v}_F\cdot(\partial_{\bf R}-2ie{\bf A})}{2\omega_n}{g}^{(0)}_{12}+\frac{[{\bf v}_F\cdot(\partial_{\bf R}-2ie{\bf A})]^2}{4(\omega_n)^2}{g}^{(0)}_{12}.\label{Ei2GL_2}
\end{equation}
The equilibrium quasiclassical $\tau_3$-Green function can be derived by Gorkov equation and its anomalous Green function is written as\cite{Silaev} 
\begin{equation}
g^{(0)}_{12}=\frac{i\Delta}{\sqrt{(\omega_n)^2+|\Delta|^2}}.
\end{equation}
Then, with the solved $g^{(0)}_{12}$ and hence $g_{12}$ in Eq.~(\ref{Ei2GL_2}), from the corresponding gap equation $\Delta=-iUN(0)\langle{g_{12}}\rangle_F$, near $T_c$, one obtains
\begin{equation}\label{GGGG}
\sum_{\omega_n>0}\Big[\frac{v_F^2(\partial_{\bf R}-2ie{\bf A})^2}{6(\omega_n)^3}-\frac{|\Delta|^2}{(\omega_n)^3}\Big]\Delta+\Big[\sum_{\omega_n>0}\frac{2}{\omega_n}-\frac{1}{UN(0)}\Big]\Delta=0.
\end{equation}
Consequently, through the mathematical calculation, the above equation becomes 
\begin{equation}
\Big[\frac{7R(3)k_F^2}{24(\pi{T})^2}\frac{(\partial_{\bf R}-2ie{\bf A})^2}{2m}-\frac{7R(3)}{8({\pi}T)^2}{|\Delta|^2}+\ln\Big(\frac{T_c}{T}\Big)\Big]\Delta=0,
\end{equation}
which exactly recovers the Ginzburg-Landau equation.

\section{Derivation of Ginzburg-Landau equation within the path-integral approach}
\label{PI2GL}

In this part, we present the derivation of Ginzburg-Landau equation within the path-integral approach at clean and stationary case.  Specifically, with vector potential alone, after the integration over Fermi field within the path-integral approach, the gap-variation related part from Eq.~(\ref{sis2}) is written as
\begin{equation} \label{U4}
S_H=-\Big[(\chi_{11}-U^{-1})\delta|\Delta|^2+\Big(\chi_{001}\delta|\Delta|\frac{e^2A^2_0v_F^2}{3}+\chi_{13}\delta|\Delta|\frac{e^2A^2}{2m}+h.c.\Big)\Big].  
\end{equation}
with
\begin{eqnarray}
 \chi_{ij}\!\!&=&\!\!\!\frac{1}{2}\sum_p{\rm Tr}[G_0(p)\tau_iG_0(p)\tau_j],  \\
  \chi_{ijk}\!\!&=&\!\!\!\sum_p{\rm Tr}[G_0(p)\tau_iG_0(p)\tau_jG_0(p)\tau_k],
\end{eqnarray}
In Eq.~(\ref{U4}), the second and third terms denote the couplings of Higgs mode to second order of light-matter interaction, i.e., second order of ${\bf k}\cdot{e{\bf A}}/m$ and linear one of $\frac{e^2A^2}{2m}\tau_3$, respectively.

Through the mathematical calculation, the coupling coefficients read
\begin{eqnarray}
  \chi_{11}\!-\!U^{-1}&=&\frac{1}{2}\sum_{p}{\rm Tr}\Big[\frac{ip_n\!+\!\xi_{\bf k}\tau_3\!+\!\Delta_0\tau_1}{(ip_n)^2-E^2_{\bf k}}\tau_1\frac{ip_n\!+\!\xi_{\bf k}\tau_3\!+\!\Delta_0\tau_1}{(ip_n)^2\!-\!E^2_{\bf k}}\tau_1\Big]\!-\!U^{-1}\!\!=\!\!\sum_p\frac{(ip_n)^2\!+\!\Delta_0^2\!-\!\xi^2_{\bf k}}{[(ip_n)^2\!-\!E^2_{\bf k}]^2}\!-\!U^{-1}\nonumber\\
  &=&\sum_p\Big\{\frac{2\Delta_0^2}{[(ip_n)^2-E^2_{\bf k}]^2}+\frac{1}{(ip_n)^2-E^2_{\bf k}}\Big\}-\sum_p\frac{1}{(ip_n)^2-E^2_{\bf k}}\nonumber\\
  &=&2{\Delta_0^2}\sum_{\bf k}\frac{1}{2E_{\bf k}^2}\Big[\frac{1\!-\!2f(E_{\bf k})}{2E_{\bf k}}\!+\!\partial_{E_{\bf k}}f(E_{\bf k})\Big]=2{\Delta_0^2}\sum_{\bf k}\frac{\partial_{E_{\bf k}}}{E_{\bf k}}\Big[\frac{2f(E_{\bf k})\!-\!1}{2E_{\bf k}}\Big],
\end{eqnarray}
\begin{eqnarray}
  \chi_{13}&=&\frac{1}{2}\sum_{p}{\rm Tr}\Big[\frac{ip_n\!+\!\xi_{\bf k}\tau_3\!+\!\Delta_0\tau_1}{(ip_n)^2-E^2_{\bf k}}\tau_1\frac{ip_n\!+\!\xi_{\bf k}\tau_3\!+\!\Delta_0\tau_1}{(ip_n)^2-E^2_{\bf k}}\tau_3\Big]=\sum_p\frac{2\xi_{\bf k}\Delta_0}{[(ip_n)^2-E^2_{\bf k}]^2}=0,\nonumber\\
    \chi_{001}&=&\sum_{p}{\rm Tr}\Big[\frac{ip_n\!+\!\xi_{\bf k}\tau_3\!+\!\Delta_0\tau_1}{(ip_n)^2-E^2_{\bf k}}\frac{ip_n\!+\!\xi_{\bf k}\tau_3\!+\!\Delta_0\tau_1}{(ip_n)^2-E^2_{\bf k}}\frac{ip_n\!+\!\xi_{\bf k}\tau_3\!+\!\Delta_0\tau_1}{(ip_n)^2-E^2_{\bf k}}\tau_1\Big]=\sum_{p}2\Delta_0\frac{E^2_{\bf k}\!+\!3(ip_n)^2}{[(ip_n)^2-E^2_{\bf k}]^3}\nonumber\\
    &=&\Delta_0\sum_{\bf k}\Big\{\Big[\Big(\frac{6}{8E_{\bf k}^3}\!-\!\frac{36E_{\bf k}}{16E_{\bf k}^4}\!+\!\frac{48E_{\bf k}^2}{32E_{\bf k}^5}\Big)f(E_{\bf k})\!+\!2\Big(\frac{6E_{\bf k}}{8E_{\bf k}^3}\!-\!\frac{12E_{\bf k}^2}{16E_{\bf k}^4}\Big)\partial_{E_{\bf k}}f(E_{\bf k})\!+\!\frac{4E_{\bf k}^2}{8E_{\bf k}^3}\partial_{E_{\bf k}}^2f(E_{\bf k})\Big]\nonumber\\
    &&\mbox{}\!+\![E_{\bf k}\rightarrow-E_{\bf k}]\Big\}=2\Delta_0\sum_{\bf k}\frac{\partial_{E_{\bf k}}^2f(E_{\bf k})}{2E_{\bf k}}. \label{ad001}
\end{eqnarray}
Then, further following the derivation of Eqs.~(\ref{bg}) and (\ref{lg}) and notation,  one has $\chi_{001}=\Delta_0\frac{7DR(3)}{2(\pi{T})^2}=2D\Delta_0\lambda_l$ and $\chi_{11}-U^{-1}=4\Delta_0^2\frac{7DR(3)}{8(\pi{T})^2}=\beta_L4\Delta_0^2D$ near $T_c$. In this circumstance, using $\Delta_0^2=-\alpha_L/\beta_L$, Eq.~(\ref{U4}) becomes
\begin{equation}\label{U7}
  S_H=-2D\Big(\alpha_L\delta|\Delta|^2+\frac{\beta_L6\Delta_0^2\delta|\Delta|^2}{2}+\lambda_L\frac{|2ieA_0|^22\Delta_0\delta|\Delta|}{4m}\Big), 
\end{equation}
which exactly recovers the nonequilibrium variation of Ginzburg-Landau Lagrangian at stationary case.  In fact, within the path-integral approach, the derivation of this Lagrangian at stationary case is exactly same as the one of effective action for Higgs mode near $T_c$ when $\Omega\rightarrow0$, guaranteeing the physical continuity between $\Omega\rightarrow0$ and $\Omega=0$.

Moreover, it is noted that mathematically,  the finite response coefficient $\chi_{001}$ in Eq.~(\ref{ad001}), which is derived at stationary situation, arises from a third-order residue, and both response coefficients $\lambda_E$ [Eq.~(\ref{DE})] derived from Eilenberger equation and $\lambda_p$ [Eq.~(\ref{lp})] derived from path-integral approach at finite optical frequency $\Omega$, for a continuous $\Omega\rightarrow0$, can exactly recover this result with $\chi_{001}=2\Delta_0\lambda_{E/p}$. But if the optical frequency $i\Omega$ in Eqs.~(\ref{DE}) and (\ref{lp}) is taken as bosonic Matsubara frequencies, one only encounters a first-order residue, and hence, finds vanishing response coefficients at all $\Omega\ne0$, leading to an unphysical abrupt change between results at $\Omega=0$ and $\Omega\rightarrow0$. ~\\

\end{appendix}

\end{widetext}


\begin{thebibliography}{0}
\bibitem{Silaev0} It is noted that Ref.~\onlinecite{Silaev} also used the diagrammatic formalism in order to justify/recover the result from the Eilenberger equation. Nevertheless, a special procedure, by first taking the commutation between the self-energy and $\tau_3$-Green function and applying the quasiclassical approximation afterwards, was applied to handle the calculation within the diagrammatic formalism (refer to Sec.~IV~A in Ref.~\onlinecite{Silaev}). This procedure is not the standard diagram formulation\cite{G1}, but follows the exact step to derive the Eilenberger equation\cite{Eilen1}. Consequently, the applied diagrammatic formalism approach with this procedure is exactly same as the Eilenberger equation rather than a separate nontrivial approach.
\end{thebibliography}
\end{document}